\documentclass[acmsmall,screen,nonacm]{acmart} 

\usepackage{booktabs}   

\usepackage{sansmathaccent}
\usepackage{amsmath}
\usepackage{pifont}
\usepackage{hhline}
\usepackage{comment}
\usepackage[caption=false]{subfig}
\pdfmapfile{+sansmathaccent.map}

\usepackage[T1]{fontenc}
\IfFileExists{luximono.sty}{\usepackage[scaled=0.9]{luximono}}{\usepackage[scaled=0.81]{beramono}}
\def\nocolour{ }

\usepackage{listings, silver}
\usepackage{color}
\usepackage{soul}
\usepackage{makecell}
\usepackage{multirow}
\usepackage{dirtree}
\usepackage{xspace}
\definecolor{light-gray}{gray}{0.9}
\definecolor{darkgreen}{rgb}{0.0,0.7,0.0}

\usepackage{tikz}

\newcommand\code[1]{{\small\texttt{#1}}}
\newcommand\secref[1]{Sect.~\ref{#1}}
\newcommand\appref[1]{App.~\ref{#1}}

\newcommand{\nref}[2]{\ref{#1}}
\newcommand\oftheTR{}
\newcommand\napprefsmall[2]{App.~\nref{#1}{#2}}
\newcommand\nappref[2]{App.~\nref{#1}{#2}}
\newcommand\ntabref[2]{Tab.~\nref{#1}{#2}}
\newcommand\nfigref[2]{Fig.~\nref{#1}{#2}}

\newcommand\defref[1]{Def.~\ref{#1}}

\newcommand\tabref[1]{Tab.~\ref{#1}}
\newcommand\figref[1]{Fig.~\ref{#1}}

\newcommand{\wrt}{{{w.r.t.\@}}}

\newcommand{\eg}{{{e.g.,~}}}
\newcommand{\ie}{{{i.e.,~}}}

\newcommand{\suchthat}{{{s.t.~}}}

\newcommand{\sound}{verification-preserving}

\newcommand{\Sound}{Verification-preserving}
\newcommand{\alwayspreserving}{always-preserving}
\newcommand{\Alwayspreserving}{Always-preserving}

\newcommand{\unsound}{non-preserving}

\newcommand{\UnSound}{Non-Preserving}
\newcommand{\soundness}{semantic}
\newcommand{\Soundness}{Semantic}
\newcommand{\longsound}{\sound}
\newcommand{\Longsound}{\Sound}

\newcommand{\SC}{SC}

\newcommand{\definedas}{\triangleq}

\reversemarginpar

\usepackage[normalem]{ulem}

\newcommand{\hoaretriple}[3]{$\{ #1 \}$ \vipercode{#2} $\{ #3 \}$}
\newcommand{\soutifcolour}[1]{}

\newcommand\todo[1]{\ifdefined\nocolour{}\else{\textcolor{red}{TODO: #1}}\fi}

\newcommand{\peter}[1]{\ifdefined\nocolour{#1}\else{\color{blue}{#1}}\fi}
\newcommand{\pout}[1]{\peter{{\soutifcolour{#1}}}}

\newcommand{\thibault}[1]{\ifdefined\nocolour{{#1}}\else{\color{blue}{#1}}\fi}
\newcommand{\tout}[1]{\thibault{\soutifcolour{#1}}}
\newcommand{\gaurav}[1]{\ifdefined\nocolour{#1}\else{\color{blue}{#1}}\fi}
\newcommand{\gout}[1]{\gaurav{{\soutifcolour{#1}}}}

\newcommand{\Viper}{\textsc{Viper}}
\newcommand{\verifast}{\textsc{VeriFast}}
\newcommand{\grasshopper}{\textsc{GRASShopper}}
\newcommand{\nagini}{\textsc{Nagini}}
\newcommand{\rslviper}{\textsc{RSL-Viper}}
\newcommand{\steel}{\textsc{Steel}}
\newcommand{\refinedc}{\textsc{RefinedC}}
\newcommand{\caper}{\textsc{Caper}}
\newcommand{\corral}{\textsc{Corral}}
\newcommand{\boogie}{\textsc{Boogie}}
\newcommand{\vercors}{\textsc{VerCors}}
\newcommand{\isabelle}{\textsc{Isabelle/HOL}}

\definecolor{darkred}{rgb}{0.55, 0.0, 0.0}
\newcommand{\tick}{\textcolor{green!75!black}{\ding{51}\xspace}}
\newcommand{\xmark}{\textcolor{darkred}{\ding{55}\xspace}}

\newcommand{\scond}[3]{\ensuremath{\mathit{SC}_{#1}^{#2}(#3)}}
\newcommand{\structcond}[3]{\ensuremath{\mathit{StructC}_{#1}^{#2}(#3)}}

\newcommand{\determ}{\ensuremath{\mathit{det}}(\varphi_1, \varphi_2, \varphi_2', s)}

\newcommand{\StructInnerUniversal}{universal determinization condition}

\newcommand{\determFraming}{\ensuremath{\mathit{det}}(\varphi, r, \varphi_r', s)}

\newcommand*{\fpointsto}[3]{\ensuremath{{#1}\overset{\scriptscriptstyle{#3}}{\mapsto}{#2}}}
\newcommand*{\pointsto}[2]{\ensuremath{{#1} \mapsto {#2}}}

\newcommand{\acq}{\texttt{acq}}

\newcommand{\Assign}[2]{\ensuremath{#1\gets#2}}

\newcommand{\ReadAccess}[2]{[#2]_{#1}}
\newcommand{\AcqReadAccess}[1]{\ReadAccess{\acq}{#1}}

\makeatletter
\def\CAS#1#2#3#4{\@ifnextchar\bgroup {\CASAssign{#1}{#2}{#3}{#4}}{\CASOp{#1}{#2}{#3}{#4}}}
\newcommand{\CASOp}[4]{\texttt{CAS}_{#1}(#2,#3,#4)}
\newcommand{\CASAssign}[5]{\Assign{#1}{\CASOp{#2}{#3}{#4}{#5}}}

\newcommand{\truesym}{\ensuremath{\textsf{true}}}

\newcommand{\whilenoinv}[2]{ \mathbf{while} \; (#1) \; \{#2\} }

\newcommand{\bwhilenoinv}{ \whilenoinv{b}{s} }

\newcommand{\NDIf}[2]{
	\mathbf{if} \; (*) \; \{#1\} \; \mathbf{else} \; \{#2\}
}

\newcommand{\bcmd}[1]{ \mathbf{#1} \; }
\newcommand{\seq}{ \mathbf{;} \; }
\newcommand{\cskip}{ \mathbf{skip} }

\newcommand{\emptyannot}{\ensuremath{\epsilon}}

\usepackage{esvect}
\newcommand{\vecto}[1]{
	\vv{\vphantom{d}#1}
}

\usepackage{algorithm}
\usepackage{algorithmic}

\usepackage{enumitem}
\setlist{nolistsep}

\renewcommand*{\proofname}{Proof sketch}
\newcommand{\smallParagraph}[1]{\vspace{1pt} \indent \textit{#1}}

\newcommand{\figurespace}{\vspace{-1mm}}

\begin{document}

\title{Verification-Preserving Inlining in Automatic Separation Logic Verifiers (extended version)}         



\author{Thibault Dardinier}
\orcid{0000-0003-2719-4856}
\email{thibault.dardinier@inf.ethz.ch}
\affiliation{
  \department{Department of Computer Science}
  \institution{ETH Zurich}
  \country{Switzerland}
}

\author{Gaurav Parthasarathy}
\orcid{0000-0002-1816-9256}
\email{gaurav.parthasarathy@inf.ethz.ch}
\affiliation{
  \department{Department of Computer Science}
  \institution{ETH Zurich}
  \country{Switzerland}
}

\author{Peter M{\"u}ller}
\orcid{0000-0001-7001-2566}
\email{peter.mueller@inf.ethz.ch}
\affiliation{
  \department{Department of Computer Science}
  \institution{ETH Zurich}
  \country{Switzerland}
}

\begin{abstract}
Bounded verification has proved useful to detect bugs and to increase confidence in the correctness of a program. In contrast to unbounded verification, reasoning about calls via (bounded) inlining and about loops via (bounded) unrolling does not require method specifications and loop invariants and, therefore, reduces the annotation overhead to the bare minimum, namely specifications of the properties to be verified.
For verifiers based on traditional program logics,
verification is preserved by inlining (and unrolling):
successful unbounded verification of a program w.r.t.\ \emph{some} annotation implies successful verification of the inlined program. That is, any error detected in the inlined program reveals a true error in the original program.
However, this essential property might not hold for \emph{automatic separation logic} verifiers such as \caper{}, \grasshopper{}, \refinedc{}, \steel{}, \verifast{}, and verifiers based on \Viper{}.
In this setting, inlining generally changes the resources owned by method executions, which may affect automatic proof search algorithms and introduce spurious errors.

In this paper, we present the first technique for \emph{\longsound{}} inlining in automatic separation logic verifiers.
We identify a semantic condition on programs and prove in \isabelle{} that it ensures \longsound{} inlining for state-of-the-art automatic separation logic verifiers. We also prove a dual result: successful verification of the inlined program ensures that there are method and loop annotations that enable the verification of the original program for bounded executions.
To check our semantic condition automatically, we present two approximations that can be checked syntactically and with a program verifier, respectively. We \thibault{implement} these checks in \Viper{} and demonstrate that they are effective for non-trivial examples from different verifiers.
\end{abstract}




\maketitle

\section{Introduction}\label{sec:introduction}
Modular deductive program verification can reason about complex programs and properties, but is expensive. Even automatic modular verifiers require a substantial annotation overhead, including method pre- and postconditions, loop invariants, and often ghost code.
Bounded verification is a powerful alternative that reduces this overhead significantly. By inlining method calls (i.e., replacing a call by the callee's body up to a finite call depth), bounded verification does not require method specifications. Similarly, it avoids the need for loop invariants by unrolling loops (i.e., replacing a loop by finitely many copies of its body).

While bounded verification generally does not prove correctness for all executions of a program, it effectively finds errors and increases the confidence that a program is correct. Consequently, bounded verification is commonly applied by model checkers~\cite{ClarkeKroeningLerda04} and also used by
deductive verifiers. For example, the \corral{} verifier~\cite{LalQL12}, which powers Microsoft's Static Driver Verifier~\cite{LalQ14}, 
inlines method calls and unrolls loop iterations in a \boogie{} program~\cite{Leino08}, before calling the deductive \boogie{} verifier.
In the following, we subsume both method inlining and loop unrolling under the term \emph{inlining}.

Inlining is also a useful stepping stone toward \emph{modular verification}, where we use ``modular verification'' to refer to unbounded verification that verifies method calls (resp.\ loops) w.r.t.\ their annotated contracts (resp.\ annotated loop invariants).
Detecting errors using inlining \emph{before} adding method specifications and loop invariants can prevent developers from wasting time attempting to annotate and verify an incorrect program.
Inlining is also useful \emph{during} the process of adding annotations to validate partial annotations before the program contains sufficient annotations to enable modular verification. For instance, inlining lets developers validate method specifications before providing loop invariants or validate partial loop invariants.

To avoid unnecessary manual labor, it is crucial that inlining itself does not introduce false positives (spurious errors):
a verification error in the inlined program should occur only if the error occurs also during the verification of the original program \emph{for all method specifications and loop invariants}.
Equivalently, if there exist annotations (possibly extending existing partial annotations) \suchthat{} the original program verifies then the inlined program must also verify. If this \tout{property }holds, we say that inlining is \emph{\sound}.
Note that even with \sound{} inlining, bounded or modular verification may report false positives if the program logic on which the verifier is based is incomplete or if valid proof obligations generated by the verifier cannot be discharged (e.g., due to limitations of SMT solvers); however, those false positives are not caused by inlining and, thus, irrelevant here.
\Longsound{} inlining ensures that errors detected by a verifier in the inlined program will cause the verifier to also reject the original program. This property increases the confidence that the original program is actually incorrect and spares developers the effort of trying to find (non-existing) annotations that make modular verification succeed.

In verifiers based on traditional program logics (like \corral{}), inlining is trivially \longsound{}. However, many \emph{automatic} verifiers based on \emph{separation logic} rely on proof search algorithms that may render inlining \emph{\unsound}.
Separation logic~\cite{Reynolds2002} (SL thereafter) uses
resources, such as
permissions to access heap locations.
These resources are owned by method executions and transferred between executions upon call and return. 
As a result, inlining a method call potentially changes the resources \tout{that are }available during the verification of the callee's method body, which may affect proof search algorithms that depend on the resources owned by a method execution, in particular, (1)~the automatic instantiation and (2)~the automatic selection of proof rules.
Both automation techniques may cause inlining to be \unsound{}, as we show in \secref{sec:problem}.

The usefulness of automatic SL verifiers (often based on SMT solvers) relies heavily on these automation techniques.
Thus, these techniques, which may cause inlining to be \unsound{}, are frequently used
(in different forms) in diverse and independently-developed verifiers such as
\caper{}~\cite{caper17} (a verifier for fine-grained concurrency), \grasshopper{}~\cite{grasshopper} (a verifier for a decidable separation logic fragment), \steel{}~\cite{FromherzRSGMMR21} (a verifier based on \textsc{F*}~\cite{fstar}),
\refinedc{}~\cite{Sammler21} (a verifier for C programs based on \textsc{Iris}~\cite{Jung2018}),
\verifast{}~\cite{Jacobs-Verifast11} (a verifier for C and Java programs),  and verifiers built on top of the \Viper{} infrastructure~\cite{MuellerSchwerhoffSummers16}
such as \nagini{}~\cite{EilersMueller18}
(a verifier for Python programs),
\rslviper{}~\cite{SummersMueller18} (a verifier for C++ weak-memory programs), and \vercors{}~\cite{BlomHuisman14} (a verifier for Java).

Our evaluation, performed on the test suites of \grasshopper{}, \nagini{}, \rslviper{}, and \verifast{}, shows that, while most method
calls (and loops) can be inlined in a \longsound{} manner, \unsound{} inlining occurs in practice in all four verifiers.
More precisely, a syntactic analysis of all files from their test suites shows that
\thibault{1053} files (out of 1562, 67\%) contain features that may result in \unsound{} inlining.
Further manual analysis of a sample of \thibault{72 files}\tout{these 1054 files} suggests that, for each verifier, between 10\% and 67\% of the sampled files contain methods (or loops)
that actually result in \unsound{} inlining for some caller context.

\paragraph{Approach.}
This paper presents the theoretical foundations for \emph{\longsound} inlining in automatic separation logic verifiers.
The core contribution is a novel semantic condition for programs that ensures that inlining is \sound, even in the presence of the automation techniques mentioned above and described in more detail in the next section.
A key virtue of this semantic condition is that it is compositional, whereas the definition of the \longsound{} property itself is not. Our semantic condition is inspired by the \emph{safety monotonicity} and \emph{framing} properties~\cite{Yang2002} of separation logics, but goes beyond those in three major ways:
(1)~We show that only a subset of statements must satisfy these properties for inlining to be \sound. (2)~Our semantic condition includes a novel monotonicity property on the final state of a statement execution.
(3)~Our semantic condition uses bounded relaxations of the properties that are weaker, but still sufficient to ensure \longsound{} inlining. All three improvements are crucial to support common use cases.

We have proved in \isabelle{} that our semantic condition is sufficient.
Since it is difficult to check directly using automatic \tout{program }verifiers,
we develop a \emph{structural condition} that approximates the semantic condition and \tout{that }can be checked using SMT-based verification tools.
We show its practicality by automating it in a tool that performs bounded verification of \Viper{} programs via \longsound{} inlining.
Errors reported by the resulting inlining feature are true errors. Our approach does not require pre- and postconditions and loop invariants, but checks partial annotations if present, which enables the use of inlining \tout{also }during the process of annotating a program. Our experiments show that the structural condition is sufficiently precise for most common use cases.

\paragraph{Contributions and outline.}
To the best of our knowledge, we present the first theoretical foundations of inlining in automatic SL verifiers. Our technical contributions are:

\begin{itemize}

\item We show why crucial automation techniques such as the automatic instantiation and the automatic selection of proof rules may cause inlining to be \unsound{} (\secref{sec:problem}).

\item We present a novel semantic condition for inlining in automatic SL verifiers.
Programs that satisfy this condition are guaranteed to be inlined in a \sound{} manner, without producing false positives. Our semantic condition takes partial annotations into account (\secref{sec:overview}).

\item  We formalize the semantic condition for a verification language that is parametrized by a separation algebra,
to capture different state models and different flavors of SL,
and prove that inlining is \longsound{} under our semantic condition in \isabelle{} (\secref{sec:soundness}).

\item We prove a dual result: inlining does not lead to false negatives other than errors that occur beyond the inlining bound (\appref{app:completeness}).

\item We define a structural condition that approximates the semantic condition, but can be checked in SMT-based program verifiers (\secref{sec:automation}).

\item We implement an inlining tool for \Viper{}, which checks the structural condition and the correctness of the inlined program (\secref{sec:evaluation}).

\item Our evaluation shows that (1) \unsound{} inlining occurs in practice,
	(2)~for many non-trivial examples, inlining is \longsound{}  and our structural condition captures this,
and (3)~\longsound{} inlining is effective in practice (\secref{sec:evaluation}).
\end{itemize}

\noindent
Our publicly-available artifact~\cite{artifact} contains 
\isabelle{} proofs of the technical results from \secref{sec:soundness},
the tool that we implemented, and the examples used in the evaluation.
%

\section{The Problem}\label{sec:problem}
In separation logic, resources (such as a permission to access a heap location) are owned by method executions, and transferred between executions upon call and return.
Thus, inlining a method call potentially changes the resources that the callee owns, which may affect proof search algorithms that depend on the resources owned by the method execution.
In this section, we show that this is the case for crucial automation techniques such as the automatic instantiation (\secref{subsec:auto-instantiation}) and the automatic selection (\secref{subsec:auto-selection}) of proof rules, by showing that both may cause inlining to be \unsound{} in several automatic SL verifiers.

\subsection{Automatic Instantiation of Proof Rules}
\label{subsec:auto-instantiation}



Applying proof rules, for instance, instantiating  quantifiers, often requires choosing a resource that is currently owned by the method. To handle a large, possibly unbounded search space, automatic SL verifiers employ heuristics for this purpose. These heuristics may behave differently when more resources are available and, thus, may make inlining \unsound.

Such heuristics are often necessary when automatic SL verifiers support \emph{imprecise assertions}. Imprecise assertions do not describe the extent of the heap precisely
(\eg because  multiple disjoint heap fragments satisfy the same assertion):
\tout{For instance, }\caper{}, \grasshopper{}, \refinedc{}, \steel{}, and \verifast{} support restricted forms of existentially-quantified assertions,
\verifast{} and \Viper{} support fractional ownership of resources~\cite{Boyland03,fractionalResources} with existentially-quantified fractions,\tout{quantification over the  fraction,}
\thibault{and \refinedc{}, \steel{}, and \Viper{}~\cite{SchwerhoffS15,DardinierWands} support magic wands.}
To prove the validity of such assertions, the majority of these verifiers use proof search algorithms (e.g., to choose a satisfying heap fragment among several suitable ones).


\peter{The effect of imprecise assertions on proof heuristics is relevant for inlining even though assertions are mainly needed for modular verification:
First, as explained in \secref{sec:introduction}, inlining is useful during the process of adding annotations to a program and, thus, must handle programs with partial annotations. Second, even bounded verification relies on method specifications for certain calls (e.g., to library methods and foreign functions). Both cases may involve imprecise assertions.}

\definecolor{royalblue}{rgb}{0.25, 0.41, 0.88}
\newcommand{\blue}[1]{{\color{royalblue}{#1}}}

\begin{figure}[t]
\begin{minipage}[t]{0.33\textwidth}
\begin{viper2}
method alloc():(l:Ref)
  requires true
  ensures Q(l)

method m(r:Ref)
{ 
  |\blue{\{ true \}}|
  a := alloc()
  |\blue{\{ Q(a) \}}|
  b := alloc()
  |\blue{\{ Q(a) * Q(b) \}}|
  n(a) 
  |\blue{\{\phantom{ }... * Q(b) \}}|
}
\end{viper2}
\end{minipage}
\hfill
\begin{minipage}[t]{0.35\textwidth}
\begin{viper2}
method crLock():(l:Ref)
  requires Q(?x)
  ensures P(l,x)

method n(a:Ref)
  // requires Q(a)
  // ensures ...
{ 
  |\blue{\{ Q(a) \}}|
  l := crLock()
  |\blue{\{ P(l,a) \}}|
  acquire(l, a)
  |\blue{\{ ... \}}|
}
\end{viper2}
\end{minipage}
\hfill
\begin{minipage}[t]{0.3\textwidth}
\begin{viper2}
method acquire(l, a:Ref)
  requires P(l,a)
  ensures ...

method m_inl() {
  |\blue{\{ true \}}|
  a := alloc()
  |\blue{\{ Q(a) \}}|
  b := alloc()
  |\blue{\{ Q(a) * Q(b) \}}|
  l := crLock()
  |\blue{\{ Q(a) * P(l,b) \}}|
  acquire(l, a) // fails
}
\end{viper2}
\end{minipage}
\figurespace{}
\caption{Example inspired by \verifast{} showing that inlining in automatic SL verifiers is potentially \unsound{} in the presence of imprecise assertions. The methods \code{alloc}, \code{crLock}, and \code{acquire} are part of a library and are specified via pre- and postconditions.
Methods \code{m} and \code{n} are the client code.
The commented-out specification of \code{n} illustrates one possible annotation with which \code{m} and \code{n} verify modularly.
\code{m\_inl} is the method \code{m} where the call to \code{n} is inlined.
\code{P} and \code{Q} denote abstract predicates. \code{Q(?x)} denotes a predicate instance with an existentially-quantified parameter \code{x}.
Proof outlines reflecting the verifier's automatically constructed proofs are shown in blue (where the proof outlines for \code{m} and \code{n} reflect modular proofs using \code{n}'s commented-out specification). The verifier fails to verify the call to \code{acquire} in \code{m\_inl}.
}
\label{fig:viper-intro-wildcard}
\end{figure}

\figref{fig:viper-intro-wildcard} illustrates the problem on an example inspired by \verifast{}.
It uses library methods \code{alloc} to create a data structure, and \code{crLock} and \code{acquire} to create and acquire a lock. The predicate \code{P} indicates that the lock is initialized; \code{Q} is the lock invariant (a real lock library would quantify over the lock invariant \code{Q}, but this aspect is irrelevant here). Inlining transforms method \code{m} into method \code{m\_inl}, in which the call to \code{n} has been replaced by its body and where we assume that the user has not given any (partial) annotation to \code{n} (yet). Note that the calls to the library methods are not inlined because those methods are annotated and may be implemented natively.

\code{crLock}'s precondition contains an existentially-quantified predicate instance \code{Q(?x)} and is, thus, imprecise. Therefore, a proof search algorithm needs to decide how to instantiate the bound variable \code{x}. During modular verification, this choice is determined by \code{n}'s precondition. For instance, the precondition \code{Q(a)} will cause the proof search to instantiate \code{x} with \code{a} since \code{Q(a)} is the only matching resource held by \code{n}. With this precondition (and a trivial postcondition), the modular proof succeeds.

However, the proof search heuristic fails for the inlined program on the right. Here, the method \code{m\_inl} owns \code{Q(a)} and \code{Q(b)} before the call to \code{crLock}. \verifast{}'s heuristic instantiates \code{x} with \code{b}. As a result, the call to  \code{acquire} fails, since the method owns \code{P(l,b)} instead of \code{P(l,a)}.

The fact that the program can be verified modularly, but fails to verify after inlining, shows that \verifast{}'s proof search heuristics make inlining \unsound;
\grasshopper{}'s, \thibault{\refinedc{}'s}, \Viper{}'s, and \steel{}'s proof search heuristics for instantiating existential quantifiers also can lead to \unsound{} inlining for similar reasons.
In all these cases, non-preserving inlining is caused by heuristics for proof automation. The inlined program is correct and could be verified manually by instantiating the quantifier with \code{a}.

\subsection{Automatic Selection of Proof Rules}
\label{subsec:auto-selection}

Many advanced separation logics support proof rules that manipulate resources in intricate ways, \gout{for instance,}\gaurav{\eg} to split and combine resources, to exchange resources, to put them under modalities, etc.
Most of these proof rules can be applied at many points in the program and proof.
To avoid exploring every possible combination, automatic SL verifiers use heuristics to decide when and how to apply the proof rules. Some of these heuristics are based on the resources currently owned by a method and may, thus, be affected by inlining, potentially making inlining \unsound.
For instance, \caper{} inspects the currently-owned resources to determine whether or not to create a shared memory region. \rslviper{} inspects the resources held by the method execution to determine the resources obtained from an atomic read operation.
Both heuristics may make inlining \unsound{}.


\begin{figure}[t]
\begin{minipage}{0.34\textwidth}
\begin{viper2}
method r(l:Ref):(a,b:Int)
  requires P(l)
  ensures A(b)
{
  |\blue{\{ P(l) \}}|
  a := read1(l)
  |\blue{\{ P(l) \}}|
  b := read2(l)
  |\blue{\{ A(b) \}}|
}
\end{viper2}
\end{minipage}
\hfill
\begin{minipage}{0.36\textwidth}
\begin{viper2}[mathescape]
method read1(l:Ref):(v:Int) 
  // requires true
  // ensures true
{ v := $\AcqReadAccess{\code{l}}$
  v := v+1 }

method read2(l:Ref):(v:Int)
  // requires P(l)
  // ensures A(v)
{ v := $\AcqReadAccess{\code{l}}$ }
\end{viper2}
\end{minipage}
\hfill
\begin{minipage}{0.3\textwidth}
\begin{viper2}
v := nondetInt();
if(perm(P(l)) >= 1) {
  exhale P(l)
  inhale A(v)
}
\end{viper2}
\end{minipage}
\figurespace{}
\caption{Example inspired by \rslviper{} showing that proof search algorithms may make inlining \unsound{}. Method \code{r}  performs two atomic read-acquire operations on the atomic memory location \code{l} and returns both values. Its specification summarizes the behavior of the code running before and after \code{r}. The precondition \code{P(l)} provides the invariant associated with location \code{l}. The postcondition \code{A(b)} indicates that the subsequent code requires the assertion that depends on the \emph{second} value that has been read. The code on the far right shows the proof strategy for $\code{v} := \AcqReadAccess{\code{l}}$, expressed in the \Viper{} intermediate language; it greedily exchanges the invariant for location \code{l} by an assertion for the read value. 
The program can be verified modularly by \rslviper{} using the commented-out specifications for \code{read1} and \code{read2}. The proof outline for \code{r} (shown in blue) reflects the corresponding proof by \rslviper{} for \code{r}.
However, \rslviper{}'s strategy makes inlining \unsound: in the inlined program, it applies the exchange to the first read operation, whereas successful verification needs to apply it to the second.
} 
\label{fig:rsl-intro-unsound}
\end{figure}

\figref{fig:rsl-intro-unsound} illustrates a simplified version of a heuristic used by 
\rslviper{}, which makes inlining \unsound.
The commented-out specifications for \code{read1} and \code{read2} just serve to illustrate successful modular verification and are not (partial) annotations provided by a user.
\rslviper{} automates the RSL logic~\cite{VafeiadisN13}, which associates an invariant---here represented by the predicate instance \code{P(l)}---with each atomic memory location \code{l}. 
The logic provides two proof rules for atomic reads (the complete RSL logic is more intricate).
In one rule, the invariant is consumed (before the location is read) and instead an assertion that depends on the read value is obtained (after the read)\gaurav{; this rule must be applied to verify \code{read2} modularly w.r.t.\ its (commented-out) annotation}.
In the other rule, no resource is consumed, and therefore the right to perform such an exchange via a (future) atomic read is retained by keeping the invariant (and thus, no assertion is obtained for the read value)\gaurav{; this rule must be applied to verify \code{read1} modularly w.r.t.\ its (commented-out) annotation}.
\rslviper{}'s proof strategy always attempts to apply the first proof rule before considering the second proof rule, that is, performs the exchange when possible. In our example, this greedy approach causes verification of the inlined program to fail because the first read consumes the invariant, such that no exchange can happen for the second read, and we do not obtain the assertion $\code{A(b)}$ for the  read value. However, the program can be verified modularly by \emph{not} passing the invariant to method \code{read1} such that the heuristic is prevented from performing the exchange for the first read. 
In~\figref{fig:rsl-intro-unsound}, the commented-out annotations for \code{read1} and \code{read2} serve to illustrate the annotation for this modular proof and the corresponding proof outline for \code{r} is shown in blue.
The fact that the program can be verified modularly while verification of the inlined program fails shows that the proof search heuristic of \rslviper{} causes inlining to be \unsound.
The inlined program is correct and could be verified manually by  applying the proof rule that exchanges the resources only for the second read operation, which demonstrates that the issue is, again, caused by heuristics for proof automation.

\rslviper{} is implemented by translating the input C++ program into the \Viper{} intermediate language. The proof search algorithm is represented explicitly in the \Viper{} program and a simplified version is shown on the right of \figref{fig:rsl-intro-unsound}. This snippet uses two dedicated statements to manipulate resources: \vipercode{inhale} obtains resources and \vipercode{exhale} releases resources. We sometimes use those operations in our examples, but no aspect of our work is specific to \vipercode{exhale} and \vipercode{inhale} operations. The same effects can be obtained, for instance, by calling a library method with a corresponding pre- or postcondition (as we do in~\figref{fig:viper-intro-wildcard}). The code snippet also uses a \emph{resource introspection expression} \vipercode{perm}. This expression yields the fractional ownership amount~\cite{Boyland03} held by the current method execution for a given resource and is used here to determine whether the resource \code{P(l)} is held by the current method execution.


\section{Semantic Condition: Key Ideas}\label{sec:overview}
In this section, we introduce the key ideas of the semantic condition that we define formally in~\secref{sec:soundness} and under which we prove that inlining is \longsound{}. 
\pout{For space reasons, t}These sections focus on \pout{methods}calls, but loops are handled \pout{in \appref{app:soundness} and \appref{sec:semantics_loops}, as well as}in the \isabelle{} formalization. From a verification point of view, loops are analogous to recursive methods, where the loop invariant acts as both the pre- and postcondition of the method. Unrolling $n$ loop iterations corresponds to inlining $n$ recursive calls.
\tout{calls to the method.}

\label{subsec:sound_inlining}

\paragraph{\Sound{} inlining}
Let $M$ be a collection of methods and let $s$ be an \emph{initial statement} that may contain calls to methods in $M$ (and no other calls). We call $(s, M)$ a program and we do not mention the tuple explicitly whenever it is clear from the context. An \emph{annotation} for $M$ consists of a pre- and postcondition for each method in $M$. A program $(s,M)$ verifies modularly w.r.t.\ an annotation $\mathcal{A}$ for $M$, if all methods in $M$ verify modularly w.r.t.\ $\mathcal{A}$ and $s$ verifies modularly w.r.t.\ to $\mathcal{A}$ (where method calls are verified using only their pre- and postconditions). 

The \emph{inlined version} of a program $(s,M)$ with bound $n$ is the statement $s$ with all \pout{method}calls substituted by their bodies up to a call stack size of $n$ (library calls may still be treated modularly). Calls that exceed the bound $n$ are replaced by
\vipercode{assume false}, such that the code afterwards verifies trivially.

Inlining is \emph{\longsound} for a program $(s,M)$ with bound $n$ if the following holds:
If the program $(s,M)$ verifies modularly w.r.t.\ \emph{some} annotation,
then the program inlined with bound $n$ also verifies.\footnote{The definition of \sound{} inlining is slightly different when the program already contains partial annotations as we discuss in \secref{subsec:partial-annotations} and  \secref{sec:soundness}.}
Consequently, if inlining is \sound{} for a program then each error in the inlined program is a \emph{true error}, i.e., corresponds to an error in the original program.

\paragraph{Semantic condition}
\label{subsubsec:semantic-condition}
  The \emph{semantic condition} (\pout{which we express formally}\peter{formalized} in \defref{def:sc}, see \secref{sec:soundness})
  is a property of a program that guarantees (but is not equivalent to) \sound{} inlining:
\pout{  In other words,} if the semantic condition holds for a statement $s$, a collection of methods $M$, and an inlining bound $n$,
  then inlining the program $(s, M)$ with bound $n$ is \sound{}.
  The semantic condition is \pout{also}parameterized by a \emph{resource bound} (a set of states), which we explain in \secref{sec:overview-resource-bound}.
  Informally, the semantic condition holds iff:
  \begin{enumerate}
    \item \thibault{entire} inlined method bodies satisfy \emph{bounded framing}, and
    \item call-free statements between\footnote{Statements before the first method call and after the last method call are also included.}
      method calls satisfy \thibault{\emph{bounded monotonicity}, \gaurav{which is} defined as the conjunction of
      \emph{bounded safety monotonicity} and \emph{bounded output monotonicity}}.
  \end{enumerate}


The rest of this section describes and illustrates the three key properties used in this definition:
\emph{framing} (\secref{subsec:framing}), \emph{safety monotonicity} (\secref{subsec:safemono}), and \emph{output monotonicity} (\secref{subsec:monotonicity}).\footnote{\peter{As we explain in this section, framing is stronger than both safety and output monotonicity; requiring these weaker properties for call-free statements is sufficient for inlining to be \sound{}.}} \secref{sec:overview-resource-bound} explains why it is sufficient to consider bounded relaxations of these properties. Finally, \secref{subsec:partial-annotations} shows how we deal with partially-annotated programs.

\paragraph{Verifier semantics}
\pout{It is important to understand that o}Our definition of \sound{} inlining
is based on the proof rules \emph{as applied by a given SL verifier}. Thus,  when we write that a program \emph{verifies}, we mean that verification \pout{actually}succeeds in a given verifier (using that verifier's proof search strategies). We refer to the proof rules as applied by a verifier as the \emph{verifier semantics} of that tool.
\thibault{For example, verification of the inlined program from \figref{fig:viper-intro-wildcard}
in a verifier that does not apply \verifast{}'s proof search heuristic could succeed and, thus, inlining could be \sound{}.}
\tout{E.g., if we verified the inlined program from \figref{fig:viper-intro-wildcard} using a more complete verifier that did not apply \verifast{}'s proof search heuristic, verification could succeed and, thus, inlining could be \sound.}It is the automation embodied in the verifier semantics of the used verifier (here, \verifast{}) that causes verification to fail.

Automatic SL verifiers track the resources held by a method execution. Thus, resources are a part of an SL verifier's state model, in addition to the program heap and store. For example, the state models of both \Viper{} and \verifast{} contain a mapping from heap locations (and predicate instances) to fractional permission amounts following the fractional permission extension of SL~\cite{Boyland03}. For heap locations, permissions are a fraction between $0$ and $1$ (a non-zero fraction permits reading, while \gout{a $1$-permission}\gaurav{the fraction $1$} permits writing).

Program operations may observe and modify the held permissions. E.g., the resource introspection expression \vipercode{perm(P(l))} on the right of~\figref{fig:rsl-intro-unsound} evaluates to the permission amount held in the verifier's state. The held permissions are modified, for instance, when an object is being allocated, via a method call that is treated modularly, or via an operation used to direct the verifier's proof search. We often model modifications of the held permissions via two dedicated statements: \vipercode{inhale A} adds the resources specified by the assertion \code{A} to the state; \vipercode{exhale A} removes these resources (following some heuristics when the assertion \code{A} is imprecise) or fails if they are not held in the current state.

\subsection{Framing}
\label{subsec:framing}

\begin{figure}[t]
\hfill
\begin{minipage}[t]{0.3\textwidth}
\begin{viper2}
method client(x:Ref)
  requires P(x)
  ensures [0.5]P(x)
        * $\thibault{\fpointsto{\code{x.f}}{\_}{0.5}}$
{ 
  callee(x) 
}
\end{viper2}
\end{minipage}
\hfill
\begin{minipage}[t]{0.31\textwidth}
\begin{viper2}
method callee(x:Ref)
  // requires [0.5]P(x)
  // ensures $\color{darkgreen}\thibault{\fpointsto{\code{x.f}}{\_}{0.5}}$
{
  |\blue{\{ [0.5]P(x) \}}|
  open P(x)
  |\blue{\{ $\fpointsto{\code{x.f}}{\_}{0.5}$ \}}|
  var v := x.f
  |\blue{\{ $\fpointsto{\code{x.f}}{\_}{0.5}$ \}}|
}
\end{viper2}
\end{minipage}
\hfill
\begin{minipage}[t]{0.37\textwidth}
\begin{viper2}
method client_inl(x:Ref)
  requires P(x)
  ensures [0.5]P(x) // fails
        * $\thibault{\fpointsto{\code{x.f}}{\_}{0.5}}$
{
  |\blue{\{ P(x) \}}|
  open P(x)
  |\blue{\{ $\pointsto{\code{x.f}}{\_}$ \}}|
  var v := x.f
  |\blue{\{ $\pointsto{\code{x.f}}{\_}$ \}}|
}
\end{viper2}
\end{minipage}
\hfill
\figurespace{}
\caption{A simple example inspired by \verifast{} showing that inlining (with bound $1$) can be \unsound{} if a method body (in this case \code{callee}'s body) is not framing.
Predicate \code{P(x)} is a predicate instance with predicate body $\pointsto{\code{x.f}}{\code{\_}}$,
which can be automatically split into two halves~\cite{fractionalResources}: \code{[0.5]P(x)} and \code{[0.5]P(x)}.
$\fpointsto{\code{x.f}}{\_}{f}$ denotes the fractional points-to assertion, representing fractional ownership amount $f$ for \code{x.f}.
\vipercode{open P(x)} is a ghost operation in \verifast{} that exchanges all the ownership of a predicate instance held by the method execution for ownership of its body,
which is needed to justify reading \code{x.f} on the next line.
Both methods \code{client} and \code{callee} verify modularly with the commented-out specification.
In the modular verification of \code{callee}, \vipercode{open P(x)} removes half ownership of \code{P(x)},
whereas it removes the full ownership in the inlined version of \code{client} (shown on the right), which is why the postcondition in
the inlined version of \code{client} does not verify.
}
\label{fig:framing}
\end{figure}

Automatic SL verifiers verify method calls modularly by releasing (exhaling) the resources specified by the callee's precondition
and then obtaining (inhaling) the resources specified by its postcondition. Resources held by the caller that are not exhaled are retained across the call, which is justified by separation logic's frame rule.
The frame rule states that if the Hoare triple $\{P\}\ s\ \{Q\}$ holds then $s$ also verifies in a larger state $P \ast R$, and the additional resources $R$ remain unchanged, that is, the triple $\{P \ast R\}\ s\ \{Q \ast R\}$ holds (provided that $s$ does not modify any variable in $R$).

However, the frame rule does not always apply to a verifier semantics, because
of the heuristics and proof search algorithms used by automatic SL verifiers. As shown in~\secref{sec:problem}, their behavior may depend on the resources held by the method execution. In particular, they may \emph{not} preserve the additional resources $R$ across an execution of $s$. As a result, even if a verifier is able to prove $\{P\}\ s\ \{Q\}$, it may \emph{not} be able to prove  $\{P \ast R\}\ s\ \{Q \ast R\}$,
which may lead to \unsound{} inlining.

To guarantee \sound{} inlining, the semantic condition requires \thibault{entire inlined} method bodies to be \emph{framing}.\footnote{In general, this requirement applies to all statements that are getting inlined\thibault{, including loops}. Since we focus exclusively on the inlining of method calls here, the call-free statements between calls need not be framing.}
Informally, a statement $s$ is \emph{framing} (in a verifier)
iff for all $P$, $Q$, and $R$, if the verifier can prove that $\{P\}\ s\ \{Q\}$ holds then it can also prove that $\{P \ast R\}\ s\ \{Q \ast R\}$ holds.
Framing is formally defined in \secref{sec:soundness} (\defref{def:framing}).

\figref{fig:framing} shows a simple \verifast{} example where inlining is \unsound{} because \code{callee}'s body
is not framing:
\verifast{} is able to prove
\hoaretriple{\code{[0.5]P(x)}}{open P(x)}{ \fpointsto{\code{x.f}}{\_}{0.5} } (in the modular case in the middle),
but not
\hoaretriple{\code{[0.5]P(x)} * \code{[0.5]P(x)} }{open P(x)}{ \fpointsto{\code{x.f}}{\_}{0.5} * \code{[0.5]P(x)} }
(in the inlined case on the right).
In other words, \verifast{} does not frame the ownership of \code{P(x)} around \vipercode{open P(x)}, since
\vipercode{open P(x)} consumes all the ownership of \code{P(x)} held by the method execution.

Both examples discussed in~\secref{sec:problem} contain method bodies that are not framing.
In \figref{fig:viper-intro-wildcard}, method \code{n}'s body is not framing, because ownership of \code{Q(x)} (for some \code{x}) might be consumed by \code{crLock()}, and thus\pout{such ownership} might not be framed around \code{crLock()} in the inlined program.
Similarly, in \figref{fig:rsl-intro-unsound}, the bodies of methods \code{read1} and \code{read2} are not framing, because
ownership of \code{P(l)} (for some \code{l}) might be consumed by $\AcqReadAccess{\code{l}}$, and thus, \pout{such ownership}might not be framed around
$\AcqReadAccess{\code{l}}$ in the inlined program.

\paragraph{Compound statements}
It is important to note that our \soundness{} condition requires the \emph{entire body} of an inlined method to be framing, but not necessarily every individual statement in the body. This difference
is crucial to capture many realistic methods that contain statements that are not framing, but nevertheless can be inlined in a \longsound{} way.
\gout{Our evaluation (\secref{sec:evaluation}) shows that many \grasshopper{}, \verifast{}, and \Viper{} programs fall into this category.}
E.g., consider a method whose body contains the following common \verifast{} pattern:
\code{\textbf{open} [?f]P(x,v); r := x.h; \textbf{close} [f]P(x,v); \textbf{return} r},
where \code{P(x,v)} is a predicate instance with predicate body $\pointsto{\code{x.h}}{\code{v}}$,
\vipercode{open} is a ghost operation that exchanges ownership of a predicate instance for ownership of its body, and \vipercode{close} performs the opposite operation.
\peter{Here}, \vipercode{open [?f]P(x,v)} exchanges a fraction \code{f} of ownership of \code{P(x,v)} for $\fpointsto{\code{x.h}}{v}{\code{f}}$,
  where \code{f} is an existentially-quantified positive fraction.
  After reading the value of \code{x.h}, \code{f} is used to restore the initial ownership of \code{P(x,v)}.
  In general, the more ownership of \code{P(x,v)} in the heap, the higher \code{f} will be instantiated by \verifast{}'s heuristic.
  Thus, \vipercode{open [?f]P(x,v)} is not framing, because ownership of \code{P(x,v)} cannot be framed around it in general.
  However, the method body as a whole is framing, and can thus can be captured by the semantic condition.

\subsection{Safety Monotonicity}
\label{subsec:safemono}

\begin{figure}[t]
\hfill
\begin{minipage}[t]{0.3\textwidth}
\begin{viper2}
method client()
{ 
  a, b := callee() 
  |\blue{\{ Q(a) \}}|
  l := crLock()
  |\blue{\{ P(l,a) \}}|
  acquire(l, a)
  |\blue{\{ ... \}}|
}
\end{viper2}
\end{minipage}
\hfill
\begin{minipage}[t]{0.3\textwidth}
\begin{viper2}[escapechar={|}]
method callee(x:Ref):
  (a,b:Ref)
  // requires true
  // ensures Q(a)
{
  a := alloc()
  |\blue{\{ Q(a) \}}|
  b := alloc()
  |\blue{\{ Q(a) * Q(b) \}}|
}
\end{viper2}
\end{minipage}
\hfill
\begin{minipage}[t]{0.38\textwidth}
\begin{viper2}
method client_inl()
{ 
  a := alloc()
  |\blue{\{ Q(a) \}}|
  b := alloc()
  |\blue{\{ Q(a) * Q(b) \}}|
  l := crLock()
  |\blue{\{ Q(a) * P(l,b) \}}|
  acquire(l, a) // fails
}
\end{viper2}
\end{minipage}
\hfill
\figurespace{}
\caption{A simplified \verifast{} example showing that inlining (with bound $1$) can be \unsound{} if a call-free statement is not safety monotonic due to \verifast{}'s heuristic for imprecise assertions.
The methods \code{alloc}, \code{crLock}, and \code{acquire} are defined in \figref{fig:viper-intro-wildcard}.
Both methods \code{client} and \code{callee} verify modularly with the commented-out specification.
In particular, because of its specification, method \code{callee} ensures only \code{Q(a)} while leaking \code{Q(b)},
and thus method \code{client} loses the ownership of \code{Q(b)} with the call to \code{callee}, which makes the call to \code{acquire(l, a)} succeed.
However, the inlined version of \code{client} (shown on the right) does not verify, since the ownership of \code{Q(b)} is not leaked.
}
\label{fig:safemono}
\end{figure}

During modular verification, each method execution starts out owning the resources described by its precondition. In contrast, at the same program point in the inlined program, the method owns all resources owned by the caller, which is a superset of those required by any precondition with which the original program verifies modularly. Thus, a statement in the inlined program will typically be verified in a state with more resources than the same statement in the original program.

To ensure that these additional resources do not lead to verification errors (and thereby to \unsound{} inlining), our \soundness{} condition requires that successful verification of a statement in some state implies successful verification in any larger state (states with more resources): Statements have to be \emph{safety monotonic}. 
A statement $s$ is safety monotonic if successful verification of $s$ starting in state $\varphi$ implies successful verification of $s$ in any \emph{larger state} $\varphi'$, i.e., if $\varphi'$ contains at least all the resources in $\varphi$ and agrees with $\varphi$ on the common resources (and variables).\footnote{\peter{Informally, in terms of Hoare triples, $s$ is safety monotonic iff, for all $P$ and $R$, if the verifier can prove $\{P\} \; s \; \{\mathit{true}\}$ 
then it can also prove  $\{P*R\} \; s \; \{\mathit{true}\}$.}}
As we will discuss in \secref{sec:related_work}, safety monotonicity has been explored in the context of separation logics, but not applied to inlining. Note that framing implies safety monotonicity.
Consequently, our \soundness{} condition requires safety monotonicity explicitly only for (potentially compound) call-free statements between calls (which includes the statement after the last method call in the initial statement).

\figref{fig:safemono} shows how a violation of safety monotonicity may lead to \unsound{} inlining.
In this simplified example, \code{l := crLock(); acquire(l, a)} is \emph{not} safety monotonic, since it verifies in a state with ownership of only \code{Q(a)}, but fails in a larger state with ownership of both \code{Q(a)} and \code{Q(b)}.

\subsection{Output Monotonicity}
\label{subsec:monotonicity}

\begin{figure}[t]
\hfill
\begin{minipage}[t]{0.3\textwidth}
\begin{viper2}
method client(x:Ref)
  requires $\pointsto{\code{x.f}}{\_}$
  ensures true
{ 
  |\blue{\{ $\pointsto{\code{x.f}}{\_}$ \}}|
  v := callee(x) 
  |\blue{\{ $\fpointsto{\code{x.f}}{\_}{0.5}$ \}}|
  if (perm(x.f) >= 1) {
    exhale $\pointsto{\code{x.f}}{\_}$
  }
  |\blue{\{ $\fpointsto{\code{x.f}}{\_}{0.5}$ \}}|
  v := callee(x) 
  |\blue{\{ \code{true} \}}|
}
\end{viper2}
\end{minipage}
\hfill
\begin{minipage}[t]{0.3\textwidth}
\begin{viper2}
method callee(x:Ref)
  : (v:Int)
  // requires $\color{darkgreen} \fpointsto{\code{x.f}}{\_}{0.5}$ 
  // ensures true
{
  v := x.f + 1
}
\end{viper2}
\end{minipage}
\hfill
\begin{minipage}[t]{0.38\textwidth}
\begin{viper2}
method client_inl(x:Ref)
  requires $\pointsto{\code{x.f}}{\_}$
  ensures true
{ 
  |\blue{\{ $\pointsto{\code{x.f}}{\_}$ \}}|
  v := x.f + 1
  |\blue{\{ $\pointsto{\code{x.f}}{\_}$ \}}|
  if (perm(x.f) >= 1) {
    exhale $\pointsto{\code{x.f}}{\_}$
  }
  |\blue{\{ \code{true} \}}|
  v := x.f + 1 // fails
}
\end{viper2}
\end{minipage}
\hfill
\figurespace{}
\caption{A simplified example showing that inlining (with bound $1$) can be \unsound{} if a call-free statement is not output monotonic.
The statement \vipercode{if (perm(x.f) >= 1) \{ exhale} $\pointsto{\code{x.f}}{\_}$ \code{\}} is safety monotonic but not output monotonic.
Both methods \code{client} and \code{callee} verify modularly with the commented-out specification.
In particular, because of its specification, method \code{callee} leaks ownership of \code{x.f}, and thus method \code{client} loses half ownership of \code{x.f} with each call to \code{callee}.
Therefore, the \code{if} branch is unreachable in the modular verification of method \code{client}.
However, the inlined version of \code{client} (shown on the right) does not verify.
Indeed no ownership of \code{x.f} is leaked in the inlined program, and thus the \code{if} branch is executed, which removes all ownership of \code{x.f}.
Verification of the line \code{v := x.f + 1} subsequently fails, because some ownership is required to read \code{x.f}'s value.
}
\label{fig:monoOut}
\end{figure}

As explained above, a statement $s$ in an inlined program is typically verified in a larger state than in the orginal program. Safety monotonocity ensures that the additional resources do not cause verification of $s$ to fail. However, they could affect the behavior of $s$ such that $s$ removes \emph{more} resources when executed in a larger state and, thereby, causes verification of subsequent statements to fail.
\figref{fig:monoOut} illustrates this problem.
Method \code{callee}'s body is framing, and the \code{if} statement in method \code{client} is safety monotonic. Nevertheless, inlining is not \sound{} because executing the \code{if} statement in a state with half ownership of \code{x.f}
leaves the state unchanged, whereas executing it in a state with full ownership of \code{x.f} results in a state with no ownership of \code{x.f}. This causes verification of the subsequent assignment to \code{v} in the inlined program to fail.

To avoid this problem, our \soundness{} condition requires statements to \peter{be \emph{output monotonic}, in addition to being safety monotonic. A statement that is both safety and output monotonic is called}
\emph{monotonic}.\footnote{\peter{Informally, in terms of Hoare triples,
$s$ is monotonic iff, for all $P$, $Q$, and $R$,
if the verifier can prove $\{P\} \; s \; \{Q\}$ then it can also prove $\{P*R\} \; s \; \{Q*\mathit{true}\}$.}}
A statement $s$ is \emph{output monotonic} if executing $s$ in a state $\varphi'$ that is larger than $\varphi$ results in a final state that is larger than 
the state obtained by executing $s$ from $\varphi$ (assuming $s$ verifies in both states).\footnote{For non-deterministic programs, one must lift the ordering to sets of states. We ignore this aspect here for simplicity, but show the lifted version in \secref{sec:soundness}.}
Output monotonicity \thibault{constrains} the \emph{effect} of a statement on subsequent statements.
It rejects statements that may remove \emph{more} resources when executed in a larger state, thereby causing the verification of subsequent statements to fail.
%
\thibault{Note that output monotonicity does not subsume safety monotonicity.
As an example, the statement \vipercode{assert perm(x.f) == 0} is not safety monotonic, since it only verifies in a state with no ownership of \vipercode{x.f}.
However, this statement is output monotonic, since it does not add or remove any resources.}
Since framing implies \thibault{both safety and output monotonicity}, our \soundness{} condition requires monotonicity explicitly only for (potentially compound) call-free statements between calls.

The \code{if} statement in \figref{fig:monoOut} is not output monotonic, which causes inlining to be \unsound. As another example, the statement \code{l := crLock()} from \figref{fig:viper-intro-wildcard} is not output monotonic.
Indeed, executing it in a state $\varphi'$ with ownership of both \code{Q(a)} and \code{Q(b)} might result in a state with ownership of both \code{Q(a)} and \code{P(l,b)},
while executing it in a state $\varphi$ with ownership of \code{Q(a)} only results in a state with ownership of \code{P(l,a)}.
While $\varphi'$ is a state with more resources than $\varphi$, the resulting states are not comparable, which violates output monotonicity.

\smallParagraph{Practical use cases.}
While any framing statement is monotonic, the converse does not hold.
In particular, a number of useful patterns are monotonic but not framing.
One example is the statement \vipercode{exhale} $\fpointsto{\code{x.f}}{\_}{\code{\textbf{perm}(x.f)}}$, which releases all ownership to \code{x.f} held by the method execution.
It is monotonic since it always verifies and the resulting state contains no permission to \code{x.f}, but it is not framing because permission to \code{x.f} cannot be framed around it.
A similar statement is used in \rslviper{} to transfer resources under a modality.
The statement \vipercode{open P} in \verifast{}, where \code{P} is a predicate, behaves similarly, as explained in \secref{subsec:framing}.
Even though the statement \vipercode{open P} is monotonic, it
is not framing, since, in general, the more ownership of \code{P} is held, the more ownership of \code{P} is exchanged, and thus ownership of \code{P} cannot be framed around this statement.
Another example of a monotonic statement that is not framing is releasing some \gout{non-zero}\gaurav{existentially-quantified} fractional ownership of a resource (wildcard in \Viper{}, dummy or existential fraction in \verifast{}), e.g., when calling a trusted library function that requires read access to a heap location.

\subsection{Bounded Relaxations}
\label{sec:overview-resource-bound}

Requiring (1)~inlined method bodies to be framing and (2)~call-free statements between calls to be monotonic
is sufficient to guarantee \sound{} inlining, but overly restrictive.
The framing and (safety and output) monotonicity properties  presented so far quantify over
\thibault{two \emph{arbitrary} states $\varphi \preceq \varphi'$.}
These properties \tout{thus }consider \emph{arbitrary} executions of a statement $s$, instead of considering what resources the inlined and original programs may actually own before executing $s$.

The example from \figref{fig:viper-intro-wildcard} illustrates why this condition is too restrictive. The statement $s \definedas ($\code{l := crLock()}$)$ is not monotonic (and thus not framing), as explained in \secref{subsec:monotonicity}. However, assume that we remove \tout{the statement }\code{b := alloc()} from the example.
In this case, inlining is \sound, since the heuristic can  instantiate \code{Q(?x)} with \code{Q(a)} only.
Nevertheless, our monotonicity requirement \tout{still }rejects this program, since $s$ is not output monotonic.
$s$ violates output monotonicity when $\varphi'$ owns both \code{Q(a)} and \code{Q(b)}
but $\varphi$ owns \code{Q(a)} only.
This violation is irrelevant in our modified example, since $s$ is only executed in states that own at most \code{Q(a)} (in both the inlined and the original program).

\paragraph{Bounded properties.}
To take into account which states can actually occur in executions of the inlined and the original program, our semantic condition requires only bounded relaxations of our framing and monotonicity properties. For output monotonicity, the bounded version is parameterized by a set of states $T$ (the \emph{resource bound}) and restricts $\varphi,\varphi'$ to be smaller than at least one state in $T$. The resource bound is set to the possible program states in the inlined program at the relevant point. The bounded relaxations of framing and safety monotonicity are analogous.

In our modified example, the statement
\vipercode{l := crLock()}
is bounded output monotonic w.r.t.\ the inlined program state $\phi_p$ before this statement, since $\phi_p$ owns \code{Q(a)} only,
and thus, $\varphi$ and $\varphi'$ over which the condition quantifies cannot own more.
In the following, when we refer to safety monotonicity, output monotonicity, or framing, we mean the bounded relaxations. 
We explain how to automatically check these properties in \secref{sec:automation}.

\paragraph{Practical use cases.}
The bounded relaxation is crucial for \verifast{} and \grasshopper{}, as we show in \secref{sec:evaluation}.
Many methods in their test suites contain existentially-quantified (and thus imprecise) assertions.
Without the relaxation, none of these methods would satisfy the \soundness{} condition, even though many of them can be inlined in a \longsound{} way in caller contexts where the existential quantifications are unambiguous, as in our modified example.

The bounded relaxation is also crucial for \nagini{}.
Python allows one to create object fields dynamically.
\nagini{} encodes the Python assignment \code{x.f = v} into \Viper{} as follows, where \code{P(x,f)} represents the permission to create the field \code{f} of \code{x}:
\vipercode{if (perm(P(x,f)) > 0) \{ exhale P(x,f); inhale} $\fpointsto{\code{x.f}}{\_}{}$ \vipercode{\}; x.f := v}.
This encoding replaces the resource \vipercode{P(x,f)}, if available, with ownership of the field.
While this encoding is not unbounded safety monotonic,
it is always bounded safety monotonic (and also framing), because \nagini{} ensures that 
$\fpointsto{\code{x.f}}{\_}{}$ and any ownership of \vipercode{P(x,f)} are mutually exclusive and hence no state in the bound $T$ contains both. Intuitively, \nagini{}'s proof search heuristics never has a choice which resource to use and, thus, cannot err.

\subsection{Inlining Partial Annotations}
\label{subsec:partial-annotations}
We conclude this section by first showing how we inline calls with partial annotations (i.e., where a subset of methods have annotations that may themselves be incomplete) and then explaining how to generalize the notion of \longsound{} inlining in the presence of partial annotations.

\paragraph{\Longsound{} inlining with partial annotations.}

As we explained in \gout{the introduction}\gaurav{\secref{sec:introduction}}, inlining is a useful stepping stone toward modular verification,
since it allows one to detect errors before \gout{completing the}\gaurav{adding} annotations and to validate partial annotations \gaurav{that arise during the iterative process of annotating methods (\eg by iteratively adding conjuncts to pre- and postconditions)}. 
\gaurav{A partial annotation is an annotation that may not yet contain enough information in order for modular verification to succeed.}
\gaurav{Verifying callees modularly with partial annotations may fail, \eg because the callee's precondition does not provide all resources needed to verify \peter{its body}.}
\gout{Partial annotations are generally
insufficient to verify method calls modularly, for instance, because a method precondition might not provide all resources needed to verify the method}
Therefore, inlining with partial annotations still reasons about calls by replacing them by the body of the callee method. Nevertheless, in order to validate partial annotations, inlining proves that they actually hold by asserting them in the inlined program. More precisely,
whenever a call to a method $m$ with a partial annotation is inlined, the inlined program \emph{asserts} $m$'s precondition, then executes $m$'s body, and finally \emph{asserts} $m$'s postcondition. Asserting the conditions checks that the resources they describe are held by the current method execution, but does not add or remove any resources.

The definition of \longsound{} inlining is adjusted accordingly in the presence of partial annotations.
Inlining a program with bound $n$ and partial annotations $\mathcal{A}$ is \longsound{} if the following holds:
If the program verifies modularly w.r.t.\ some \gout{(complete)}annotation \emph{that is \thibault{more complete} than} $\mathcal{A}$ (i.e., all method pre- and postconditions are \tout{logically }stronger than the corresponding pre- and postconditions in $\mathcal{A}$), then the program inlined with bound $n$ and partial annotations $\mathcal{A}$ verifies.
\gout{With this definition, true errors detected by \sound{} inlining include violations of existing partial annotations.}
\gaurav{Thus, if inlining with $\mathcal{A}$ is \sound{} for a program, then an error in the inlined program implies that the original program cannot be verified modularly for any annotation that is more complete than $\mathcal{A}$ (\eg no conjuncts can be added to $\mathcal{A}$ to make the original program verify).}

We also adjust the \soundness{} condition to take partial annotations into account, by first applying a syntactic transformation on the program that asserts the partial annotations before and after method calls (we make this precise in the next section).

\begin{figure}[t]
\begin{minipage}{0.38\textwidth}
\begin{viper2}
method b(x,y:Ref) 
  requires $\code{P(x,v)} * \pointsto{\code{y.h}}{\_}$
  ensures $\code{P(x,v)} * \pointsto{\code{y.h}}{\code{v}}$
{ c(x,y) }
\end{viper2}
\end{minipage}
\hfill
\begin{minipage}{0.32\textwidth}
\begin{viper2}[escapechar={|}]
method c(x,y:Ref)
  |\color{blue} \textbf{requires} [?f]P(x,v)|
  |\color{blue} \textbf{ensures} [f]P(x,v)|
{ |\color{darkred} \textbf{open} [f]P(x,v)|
  y.h := x.h + 1
  |\color{darkred} \textbf{close} [f]P(x,v)| }
\end{viper2}
\end{minipage}
\begin{minipage}{0.3\textwidth}
\begin{viper2}[escapechar={|}]
  |\color{blue} \textbf{assert} [?f]P(x,v)|
  |\color{darkred} \textbf{open} [f]P(x,v)|
  y.h := x.h + 1 
  |\color{darkred} \textbf{close} [f]P(x,v)|
  |\color{blue} \textbf{assert} [f]P(x,v)|
\end{viper2}
\end{minipage}
\figurespace{}
\caption{A \verifast{} example showing \emph{\sound{}} inlining in the presence of ghost code and partial annotations. \code{c}'s ghost code (in red) requires \code{c}'s partial specification (in blue) to bind the existential parameter \code{f}.
\thibault{Predicate \code{P(x,v)} is a predicate instance with predicate body \pointsto{\code{x.h}}{\code{v}}.}
The snippet on the very right shows the inlined body of \code{b} when \code{c}'s specification and ghost code are included.
} 
\label{fig:ghost-intro}
\end{figure}

\paragraph{Example.}
Consider the \tout{\verifast{} }example \thibault{\gaurav{showing} \sound{} inlining} in \figref{fig:ghost-intro}, which includes a partial annotation for method \code{c} and \tout{also }ghost code in \code{c}'s method body.
This example shows a scenario where (1)~\sound{} inlining can be used to find errors with partial annotations, and (2)~inlining the method body makes sense only if one takes partial annotations into account. By asserting partial annotations in the inlined program, our technique handles both aspects.

This example is based on the common \verifast{} pattern already described in \secref{subsec:framing}.
The existential quantification over \code{f} in \code{c}'s specification  enables more possible callers and transfers back the initial ownership.
The ghost operations \vipercode{open [f]P(x,v)} and \vipercode{close [f]P(x,v)} are required, 
since the verifier does not automatically unroll \code{P(x,v)} to justify reading \code{x.h} (the predicate body of \code{P(x,v)} is \pointsto{\code{x.h}}{\code{v}}).
These operations have a meaning only due to \code{c}'s precondition that binds \code{f},
which shows that we need to inline partial annotations.
Note that \code{c}'s specification is truly partial, since ownership of \code{y.h} would be required to justify the assignment in a modular proof.

The inlined body of \code{b} with its specification and ghost code is shown on the right of~\figref{fig:ghost-intro}.
Asserting the precondition \code{[?f]P(x,v)} checks whether some ownership of \code{P(x,v)} is held and \code{f} binds the fractional ownership amount that \verifast{} picks to prove the assertion.
In this case, \verifast{}'s heuristic binds \code{f} to the currently-owned fraction of \code{P(x,v)}, i.e., to $1$.

The inlined program fails, since \code{y.h} does not hold the same value as \code{x.h}, which is required by \code{b}'s postcondition.
Since there is no specification for \code{c} that can make \code{b} verify, inlining is \longsound{} and thus, inlining detects a true error without the user having to provide ownership of \code{y.h} in \code{c}'s specification.
The \soundness{} condition holds in this example, because the inlined body of \code{c} (including the \vipercode{assert} statements) \emph{as a whole} satisfies the frame rule.

\section{Verification-Preserving Inlining} \label{sec:soundness}
In \tout{the previous section}\secref{sec:overview}, we motivated the building blocks of the \soundness{} condition.
In this section, we \tout{give a formal definition of}\thibault{formally define} the \soundness{} condition and prove that inlining is \sound{}
when the \soundness{} condition holds.
In order to express this formal result in a general way, we define a parametric verification language that captures the essence of verification languages such as \grasshopper{}, \verifast{}, and \Viper{}.
To capture different models of resources, the states of this language are elements of a separation algebra.
We formalize inlining and the \soundness{} condition for this language, and prove that inlining is \longsound{} under the \soundness{} condition (Theorem~\ref{thm:soundness}).
We first consider a version of inlining that ignores annotations, and then show how to leverage this version to support inlining with partial annotations.
All results presented in this section have been mechanized in \isabelle{}~\cite{artifact}.
As explained in \secref{sec:overview}, this section focuses on methods calls, but loops are handled in \napprefsmall{app:soundness}{A} and
\napprefsmall{sec:semantics_loops}{B}\oftheTR{}, and in the mechanization.

\subsection{\thibault{State Model and Verification Language}}
\label{subsec:language}


We present the essential aspects of our verification language here; additional formal definitions are given in \nappref{app:soundness}{A}.

\paragraph{State model}
To reflect the \emph{verifier semantics} (see \secref{subsec:sound_inlining}), our state model contains separation logic resources in addition to a standard state with local variables and a heap. The models of verifiers such as \grasshopper{}, \verifast{}, or \Viper{} are essentially captured by a separation algebra~\cite{Calcagno2007,Dockins2009} where $\Sigma$ is the set of states,
$\oplus: \Sigma \times \Sigma \rightarrow \Sigma$ is a partial operation that is commutative and associative,
and $u \in \Sigma$ is a neutral element for $\oplus$.
Intuitively, two states can be added if they agree on the values of common local variables and heap locations and if their resources can be combined in a consistent way. The addition then contains the union of their local variables and heap values, and the combination of their resources.
We write $\varphi \# \varphi'$ if $\varphi \oplus \varphi'$ is defined, and $\varphi' \preceq \varphi \Longleftrightarrow (\exists \varphi'' \in \Sigma.\; \varphi = \varphi' \oplus \varphi'')$.
We lift the operators $\oplus$ and $\preceq$ to sets of states $T$ and $U$, where
$T \oplus U \definedas{} \{ \varphi \oplus \varphi' | \varphi \in T \land \varphi' \in U \land \varphi \# \varphi' \}$ and
$U \preceq T \Longleftrightarrow (\forall \varphi \in T.\; \exists \varphi' \in U.\; \varphi' \preceq \varphi)$.

\paragraph{Language and semantics}
We consider a parametric verification language with the previously-described state model and the following statements:
{
	\setlength{\abovedisplayskip}{5pt}
	\setlength{\belowdisplayskip}{5pt}
	\setlength{\abovedisplayshortskip}{0pt}
	\setlength{\belowdisplayshortskip}{0pt}
	\begin{align*}
		S
		\thibault{\Coloneqq}
		\; &S \seq S \; | \; \NDIf{S}{S} 
		\; | \; \whilenoinv{A}{S}
		\; | \; \vecto{V} := m(\vecto{V})
                \; | \; \cskip\\
		\; &| \; \bcmd{assume} A
		\; | \; \bcmd{assert} A
		\; | \; \bcmd{inhale} A 
		\; | \; \bcmd{exhale} A
		\; | \; \bcmd{var} \vecto{V} \; | \; \bcmd{havoc} \vecto{V}
		\; | \; \bcmd{custom} O 
	\end{align*}
}
\noindent
where $A$ represents assertions and $\vecto{V}$ lists of variable identifiers. $O$ is a parameter of the language used to represent verifier-specific statements, such as \vipercode{open} and \vipercode{close} in \verifast{}.

Most non-custom statements of this language are standard and have the usual semantics. Our \code{if} statement is non-deterministic and can model both non-deterministic choice and (by using $\bcmd{assume}$ statements in the branches) deterministic branching. As explained earlier, $\bcmd{inhale} A$ combines the current verification state with a state satisfying $A$, while $\bcmd{exhale} A$ removes a state satisfying $A$ from the current verification state (and fails if this is not possible). 
During verification of $\bcmd{inhale} A$, the verifier must consider \emph{all possible} states that satisfy $A$. However, for $\bcmd{exhale} A$ it can \emph{choose} how to satisfy $A$, for instance, how to instantiate an existential quantifier. This choice is embodied by a heuristic, which is a parameter of our verifier semantics.
The only assumptions we make about the verifier's heuristics is that they are \emph{local} and \emph{deterministic}, \ie{} the choices made by the verifier are fully determined by
the current verification state. For instance, for the same verification state, a heuristic will always make the same choice for an existential quantifier.
This is the case for verifiers such as \grasshopper{}, \verifast{}, and verifiers built on top of \Viper{}.
In contrast, \caper{} uses backtracking, which is not local but depends on the whole program.

For an annotation $\mathcal{A}$ containing pre- and postconditions for every method (transitively) called from the statement $s$, we write $ver_\mathcal{A}(\varphi, s)$ if\gout{the statement} $s$ verifies modularly for executions starting in the initial state $\varphi$, where
method calls are verified using only their pre- and postconditions in $\mathcal{A}$.
In this case, we define $sem_\mathcal{A}(\varphi, s)$ as the set of states that are reached after executing $s$ in the state $\varphi$ w.r.t. $\mathcal{A}$.
Note that $ver_\mathcal{A}(\varphi, s)$ and $sem_\mathcal{A}(\varphi, s)$ model the verifier semantics (as introduced in~\secref{subsec:sound_inlining}).

\label{sec:inlining-soundness}

\subsection{Inlining without Annotations}
\label{subsec:inlining-no-annotations}

We now formally define inlining and the \soundness{} condition for our language, and then prove that the latter implies \longsound{} inlining.
We
ignore annotations here, but they are handled in \secref{sub:inl-partial-annot}.
While inlining, we need to keep track of the already-used variables, to avoid variable capturing.
For simplicity, we ignore all renaming issues here, but our \isabelle{} formalization covers this aspect.
The inlining function 
$\mathit{inl}_M^n(s)$ yields the statement $s$ where all calls to methods from $M$ are substituted by their bodies up to the inlining bound $n$ (annotations are ignored):


\begin{definition}\label{def:inline_function}\textbf{Inlining (ignoring renaming issues and loops).}
	\begin{align*}
		\mathit{inl}^n_M(s) &\definedas s \text{ (if $s$ is call-free)} &
		\mathit{inl}^0_M(\vecto{y} \coloneqq m(\vecto{x})) &\definedas \bcmd{assume} \bot \\
		\mathit{inl}^n_M(s_1 \seq s_2) &\definedas \mathit{inl}^n_M(s_1) \seq \mathit{inl}^n_M(s_2) &
		\mathit{inl}^{n+1}_M(\vecto{y} \coloneqq m(\vecto{x})) &\definedas \mathit{inl}^n_M(s_m) \\
		\mathit{inl}^n_M(\NDIf{s_1}{s_2}) &\definedas \NDIf{\mathit{inl}^n_M(s_1)}{\mathit{inl}^n_M(s_2)}
	\end{align*}
\noindent where $s_m$ is the body of method $m \in M$ \thibault{with arguments correctly substituted}.

\end{definition}

When the \tout{remaining }inlining bound $n$ has reached $0$, additional calls render the execution infeasible.
Otherwise, a method call is replaced by the method body (with suitable substitutions, omitted here).
%

\paragraph{Monotonicity and framing}
Before we show the semantic condition, we formalize its key building blocks.
\thibault{To specify the bounded relaxation (motivated in~\secref{sec:overview-resource-bound}),
we define the shorthand $(\varphi \ll T) \definedas (\exists \varphi' \in T \ldotp \varphi \preceq \varphi')$.}
The following definition combines the bounded safety and output monotonicity properties motivated in~\secref{subsec:safemono} and~\secref{subsec:monotonicity}:

\begin{definition}\textbf{Bounded safety and output monotonicity.}\label{def:mono}
	\begin{align*}
		\mathit{mono}_\mathcal{A}(T, s) \thibault{\triangleq} &(\forall \varphi_1, \varphi_2 \in \Sigma \ldotp
		\varphi_1 \preceq \varphi_2 \ll T
		\land \mathit{ver}_\mathcal{A}(\varphi_1, s) \\
		&\Longrightarrow
		\mathit{ver}_\mathcal{A}(\varphi_2, s) \land \phantom{} \tag{safety monotonicity} \\
		&\quad \quad
		\mathit{sem}_\mathcal{A}(\varphi_1, s) \preceq \mathit{sem}_\mathcal{A}(\varphi_2, s)) \tag{output monotonicity}
	\end{align*}
\end{definition}

\noindent
$\mathit{mono}_\mathcal{A}(T, s)$ states that
if \tout{the statement }$s$ verifies in a state $\varphi_1$, then $s$ also verifies in \peter{a larger state $\varphi_2$} (safety monotonicity), and \tout{the execution of}\thibault{executing} $s$ in $\varphi_2$ results in a larger set of states than \tout{the execution}\thibault{executing $s$}
in $\varphi_1$ (output monotonicity).
\thibault{$\varphi_2 \ll T$ expresses that the} \peter{larger state $\varphi_2$ is smaller than at least one} state of $T$.

The following definition captures the bounded framing property motivated in~\secref{subsec:framing}:
\begin{definition}\textbf{Bounded framing.}\footnote{For the sake of presentation, we ignore the (usual) side condition that $r$ does not contain variables modified by the statement $s$, but this is handled in our \isabelle{} formalization.}
	\label{def:framing}
	\begin{align*}
		\mathit{framing}_\mathcal{A}(T, s) \thibault{\triangleq}
		(
		&\forall \varphi, r \in \Sigma.\; \varphi \# r \land
		\thibault{ \varphi \oplus r \ll T }
		\land \phantom{}
		\mathit{ver}_\mathcal{A}(\varphi, s) \\
		&\Longrightarrow
		\mathit{ver}_\mathcal{A}(\varphi \oplus r, s) \land
		\mathit{sem}_\mathcal{A}(\varphi, s) \oplus \{ r \}
		\preceq
		\mathit{sem}_\mathcal{A}(\varphi \oplus r, s)
		)
	\end{align*}
\end{definition}

\noindent
$\mathit{framing}_\mathcal{A}(T, s)$ holds iff,
for any state (smaller than some state in $T$) that can be decomposed into $\varphi \oplus r$ s.t.\ 
$s$ verifies in $\varphi$, 
it holds that executing $s$ in the state $\varphi \oplus r$ verifies
and results in a larger\footnote{\thibault{It would also be correct to require
$\mathit{sem}_\mathcal{A}(\varphi \oplus r, s) = \mathit{sem}_\mathcal{A}(\varphi, s) \oplus \{ r \}$
instead of $
\mathit{sem}_\mathcal{A}(\varphi, s) \oplus \{ r \}
\preceq
\mathit{sem}_\mathcal{A}(\varphi \oplus r, s)
$.
However, there are cases where having more resources available leads to more resources being generated by a statement
(\eg changing the modality of a resource in RSL-Viper), and these cases are captured only by the latter (weaker) requirement.}} set of states
than executing $s$ in $\varphi$ and adding the frame $r$ afterwards
(recall that we have lifted the $\oplus$ and $\preceq$ operators to sets of states).
As an example, the statement \vipercode{exhale\ $\fpointsto{\code{x.f}}{\_}{\code{\textbf{perm}(x.f)}}$} is framing if and only if no state in $T$ contains non-zero ownership of \vipercode{x.f}.
Otherwise, we can prove that it is not framing, by choosing a frame $r$ with non-zero ownership of \vipercode{x.f}.

\paragraph{Semantic condition}
For an inlining bound $n$, a set of methods $M$, a set of states $T$, and a statement $s$,
we denote our \soundness{} condition by $\SC{}_M^n(T, s)$,
and define it as follows (where the set of states $T$ is the bound we use for mono and framing):

\begin{definition}\textbf{\Soundness{} condition (ignoring renaming issues and loops).}
	\label{def:sc}
	\begin{align*}
		\SC{}^n_M(T, s) &\thibault{\triangleq} \mathit{mono}_{\emptyannot{}}(T, s) \tag{if $s$ is call-free} \\
		\SC{}_M^n(T, s_1 \seq s_2) &\thibault{\triangleq} \SC{}^n_M(T, s_1) \land \SC{}^n_M(\overline{\mathit{sem}}_{\emptyannot}(T, \mathit{inl}_M^n(s_1)), s_2) \\
		\SC{}_M^n(T, \NDIf{s_1}{s_2}) &\thibault{\triangleq} \SC{}^n_M(T, s_1) \land \SC{}^n_M(T, s_2) \\
		\SC{}_M^0(T, \vecto{y} := m(\vecto{x})) &\thibault{\triangleq} \top \\
		\SC{}_M^{n+1}(T, \vecto{y} := m(\vecto{x})) &\thibault{\triangleq}
		\mathit{framing}_{\emptyannot{}}(T, \mathit{inl}_M^n(s_m))
			\land \SC{}^n_M(T, s_m)
 	\end{align*}
	 where $\emptyannot{}$ is the empty annotation, $s_m$ is the body of method $m \in M$ \thibault{with arguments correctly substituted}, and
	$\overline{\mathit{sem}}_{\mathcal{A}}(T, s) \thibault{\triangleq} \bigcup_{\varphi \in \Sigma | (\exists \varphi' \in T. \varphi \preceq \varphi') \land \mathit{ver}_{\mathcal{A}}(\varphi, s)}( \mathit{sem}_{\mathcal{A}}(\varphi, s) )$.
	\\
\end{definition}

\noindent
As explained in \secref{sec:overview}, we require call-free statements to be mono, and inlined method bodies to be framing.
For the sequential composition, we need the auxiliary function $\overline{sem}$,
which applies the $\mathit{sem}$ function to all states $\varphi$ in which $s$ verifies and that is smaller than some state in T.
This auxiliary function is required to compute the right resource bound for the framing and monotonicity properties, in order to ensure \longsound{} inlining.
Note that the \soundness{} condition does not depend on any annotation, since mono and framing are enforced only on call-free statements.

\paragraph{\Sound{} inlining}
Using inlining and the \soundness{} condition, we can express and prove the following theorem (under some standard
well-formedness conditions).

\begin{theorem}\textbf{\Longsound{} inlining.}
	For any well-formed program $(s, M)$ for which
	$\SC{}_M^n(\{u\}, s)$\footnote{Recall that $u$ is the neutral element of $\oplus$, that is, the empty state.} holds,
	if there exists an annotation $\mathcal{A}$ for $M$ such that:
	\begin{enumerate}
		\item all methods in $M$ verify modularly w.r.t.\ $\mathcal{A}$, and
		\item the initial statement $s$ verifies modularly w.r.t.\ $\mathcal{A}$, that is, $\mathit{ver}_\mathcal{A}(u, s)$,
	\end{enumerate}
	then the program $(s, M)$ inlined with the inlining bound $n$ verifies:
	$\mathit{ver}_{\emptyannot{}}(u, \mathit{inl}_M^n(s))$.
	\label{thm:soundness}
\end{theorem}

\begin{proof}	
%
\thibault{\peter{We prove the following invariant relating the original and the inlined program} (assuming the \soundness{} condition holds and the original program verifies modularly):
The verification state of the inlined program has at least as many resources as the corresponding verification state during modular verification of the original program.
Formally, we prove that, for any two states $\varphi \preceq \varphi'$ smaller than
some
state in the
inlined program at the relevant point (\ie the resource bound),
if $\mathit{ver}_\mathcal{A}(\varphi, s)$ holds, then
$\mathit{ver}_{\emptyannot{}}(\varphi', \mathit{inl}_M^n(s))$ and
$
\mathit{sem}_{\mathcal{A}}(\varphi, s)
\preceq
\mathit{sem}_{\emptyannot{}}(\varphi', \mathit{inl}_M^n(s))
$ hold.
We prove this invariant to hold before and after every method call (and every loop iteration), by induction on the structure of the inlined program. In the case where we inline a method call, we know that this method call has been modularly verified using the frame rule.
We use the fact that the inlined method body is framing (from the semantic condition) to mimic the application of the frame rule for the inlined program.
In the case of a call-free program statement, we use the fact that this call-free statement is monotonic to prove that \peter{it preserves the aforementioned invariant}.}
\tout{The full proof is included in our \isabelle{} formalization~\cite{artifact}.}
\end{proof}

Since inlining and the \soundness{} condition do not depend on any annotation, this theorem implies the following result, which we are mostly interested in and have proved in \isabelle{}:
If the verification of the inlined program fails and the \soundness{} condition holds, then there does not exist an annotation such that the original program verifies modularly w.r.t.\ this annotation.
In other words, any error in the inlined program corresponds to a true error in the original program.

\subsection{Inlining with Partial Annotations}
\label{sub:inl-partial-annot}

We extend the formalization to handle partially-annotated programs in two steps:
We first apply a syntactic transformation $\mathit{assertAnnot}$  on the program that adds $\mathbf{assert}$ statements to check method specifications,
and then inline the resulting program using the annotation-agnostic $\mathit{inl}$ function defined in~\secref{subsec:inlining-no-annotations}.
This two-step approach allows us to leverage our previous results to prove that the \soundness{} condition guarantees that inlining with partial annotations is \longsound{} (Theorem~\ref{thm:extended-soundness}),
which we have also formalized and proved in \isabelle{}~\cite{artifact}.

\begin{definition}\textbf{The $\mathit{assertAnnot}$ syntactic transformation.}\\
	Let $(s,M)$ be a program with an annotation $\mathcal{A}$.
	$\mathit{assertAnnot}_\mathcal{A}(s,M)$ returns the program\\
	$(\mathit{assertAnnotStmt}_\mathcal{A}(s), \mathit{assertAnnotMethods}_\mathcal{A}(M))$.
	$\mathit{assertAnnotStmt}_\mathcal{A}(s)$ asserts the method precondition (resp.\ postcondition) before (resp.\ after) each method call in $s$.
	In particular,\\
	$\mathit{assertAnnotStmt}_\mathcal{A}(\vecto{y} := m(\vecto{x})) \definedas \bcmd{assert} P * \truesym{}$; $\vecto{y} := m(\vecto{x}); \bcmd{assert} Q * \truesym{}$
	where $P$ (resp.\ $Q$) is method $m$'s precondition (resp.\ postcondition) in $\mathcal{A}$ with the arguments correctly substituted.\\
	$\mathit{assertAnnotMethods}(M)$ returns the same methods as $M$, but where the method body $s_m$ of $m \in M$ is modified to
check the pre- and postcondition of $m$ and all methods it calls:
	$\bcmd{assert} P * \truesym{}$; $\mathit{assertAnnotStmt}_M(s_m); \bcmd{assert} Q * \truesym{}$,
	where $P$ (resp.\ $Q$) is $m$'s precondition (resp.\ postcondition) in $\mathcal{A}$ with the arguments correctly substituted.
	\label{def:assertAnnot}
\end{definition}

Given a program $(s, M)$, a bound $n$, and an annotation $\mathcal{A}$,
we define its inlined version \emph{with partial annotations} as
$\mathit{inl}_{M_\mathcal{A}}^{n}(s_\mathcal{A})$ where $(s_\mathcal{A},M_\mathcal{A}) = \mathit{assertAnnot}_\mathcal{A}(s,M)$.
For each call, the resulting inlined program asserts the precondition right before the call and also at the beginning of the callee (analogously for the postcondition).
While it seems redundant to assert the precondition twice in the inlined program, the second assertion right at the beginning of the callee makes the \soundness{} condition defined in~\secref{subsec:inlining-no-annotations} more precise: The assertion forces properties on the inlined method body to only take into account states at the beginning of the body that satisfy the precondition (and analogously for the postcondition).
Note that conjoining the precondition (resp.\ postcondition) with $\truesym{}$ in $\mathit{assertAnnot}$ is crucial for verifiers based on classical SL (such as \grasshopper{}), because for calls one must check that the caller context has \emph{at least} the resources specified by the precondition (resp.\ postcondition) before the call (resp.\ after the call).

We can now state and prove the following theorem:

\begin{theorem}\textbf{\Longsound{} inlining with partial annotations.}\\
	Let $(s, M)$ be a well-formed program with annotations $\mathcal{A}$ and $\mathcal{B}$ s.t.\ 
	$\mathcal{B}$ is \thibault{more complete} than $\mathcal{A}$ (i.e., 
	all method pre- and postconditions in $\mathcal{B}$ are logically stronger than the corresponding pre- and postconditions in $\mathcal{A}$),
	and let $(s_\mathcal{A}, M_\mathcal{A}) = \mathit{assertAnnot}_\mathcal{A}(s, M)$.
	If
	\begin{enumerate}
		\item $\SC{}_{M_\mathcal{A}}(\{u\}, s_\mathcal{A})$ holds, and
		\item all methods in $M$ verify modularly w.r.t.\ $\mathcal{B}$, and
		\item the initial statement $s$ verifies modularly w.r.t.\ $\mathcal{B}$ that is, $\mathit{ver}_{\mathcal{B}}(u, s)$,
	\end{enumerate}
	then the program $(s_\mathcal{A}, M_\mathcal{A})$ inlined with annotation $\mathcal{A}$ and inlining bound $n$ verifies:
	$\mathit{ver}_{\emptyannot{}}(u, \mathit{inl}_{M_\mathcal{A}}^n(s_\mathcal{A}))$.
	\label{thm:extended-soundness}
\end{theorem}

\begin{proof}
\thibault{
Let
$(s_\mathcal{B}, M_\mathcal{B}) \triangleq \mathit{assertAnnot}_\mathcal{B}(s, M)$.
Using (2) and (3), we prove that $(s_\mathcal{B}, M_\mathcal{B})$ verifies modularly \wrt{} $\mathcal{B}$:
This holds because the additional assertions in $(s_\mathcal{B}, M_\mathcal{B})$ reflect what must hold before and after each call when modularly verifying \wrt{} $\mathcal{B}$.
Moreover, using (1) and the fact that $\mathcal{B}$ is more complete than $\mathcal{A}$,
we prove that $\SC{}_{M_\mathcal{B}}(\{u\}, s_\mathcal{B})$ holds
(using that $(s_\mathcal{A}, M_\mathcal{A})$ and $(s_\mathcal{B}, M_\mathcal{B})$ differ only in their \vipercode{assert} statements).
We can now apply Theorem~\ref{thm:soundness} to 
$(s_\mathcal{B}, M_\mathcal{B})$ with annotation $\mathcal{B}$,
which gives us $\mathit{ver}_\epsilon(u, \mathit{inl}_{M_\mathcal{B}}^n(s_\mathcal{B}))$.
This implies $\mathit{ver}_\epsilon(u, \mathit{inl}_{M_\mathcal{A}}^n(s_\mathcal{A}))$,
because the statements $\mathit{inl}_{M_\mathcal{B}}^n(s_\mathcal{B})$
and
$\mathit{inl}_{M_\mathcal{A}}^n(s_\mathcal{A}))$
differ only in their \vipercode{assert} statements, and, since $\mathcal{B}$ is more complete than $\mathcal{A}$,
successful verification of the \vipercode{assert} statements in $\mathit{inl}_{M_\mathcal{B}}^n(s_\mathcal{B})$
implies 
successful verification of the \vipercode{assert} statements in $\mathit{inl}_{M_\mathcal{A}}^n(s_\mathcal{A})$.
\tout{The full proof is included in our \isabelle{} formalization~\cite{artifact}.}}
\end{proof}

Similarly to Theorem~\ref{thm:soundness}, we are particularly interested in the following corollary, which we have proved in \isabelle{}:
If $\SC{}_{M_\mathcal{A}}(\{u\}, s_\mathcal{A})$ holds, and verification of $\mathit{inl}_{M_\mathcal{A}}^n(s_\mathcal{A})$ fails,
then there does not exist an annotation $\mathcal{B}$ \thibault{more complete} than $\mathcal{A}$ such that
$(s, M)$ verifies modularly w.r.t.\ $\mathcal{B}$.
That is, there is no way to \thibault{complete} the partial annotation $\mathcal{A}$ \thibault{(\gaurav{\eg by adding} conjuncts to pre- or postconditions)} such that the program verifies modularly.


\section{Automation for Verification-Preserving Inlining}\label{sec:automation}
Theorems~\ref{thm:soundness} and~\ref{thm:extended-soundness}  from \secref{sec:soundness} state that errors in the inlined program correspond to true errors in the original program, \emph{provided} that the \soundness{} condition holds for this program and the inlining bound.
While inlining (\defref{def:inline_function})
and the syntactic transformation (\defref{def:assertAnnot}) are straightforward to implement, checking the \soundness{} condition directly is challenging. Both the mono and framing properties \pout{of the \soundness{} condition }are hyperproperties~\cite{hyperproperties} (properties of multiple executions) that combine universal and existential quantification over states.
Automatic program verifiers can check properties \emph{for all} executions, but \peter{cannot reason} about the existence of executions. To work around this limitation, we present two conservative approximations of the \soundness{} condition that can be checked syntactically and with a standard program verifier, respectively. 

\paragraph{Syntactic condition}
We first provide \emph{syntactic} versions of mono and framing that are fast and easy to check, to quickly accept programs that do not use features that could lead to \unsound{} inlining.
A program is syntactic mono (resp. framing) if the program does not contain any \emph{syntactic} features that could be the reason for violating mono (resp. framing). Such \emph{violating features} include operations that inspect the resources held in a state (\eg{} the \vipercode{perm} feature in \Viper{}).
Violating features also include any feature that could trigger proof search strategies for imprecise assertions.
In \grasshopper{}, \verifast{}, and \Viper{}, this includes\gaurav{,} for instance\gaurav{,} imprecise assertions in preconditions of library methods.
The syntactic check overapproximates the imprecise assertions by checking for components such as existentially-quantified parameters that could be the cause of imprecision. In~\nappref{app:syntactic-cond}{E}, we provide details about violating features for the three verifiers.

These syntactic checks are useful to quickly identify programs for which inlining is clearly \sound{}\tout{. However, they}\thibault{, but} are
too coarse to validate non-trivial applications of advanced features (including the examples in~\secref{sec:problem}).
For example, the statement on the left of \figref{fig:instrumentation} is mono and framing,
but is rejected by the syntactic check since it uses \Viper{}'s \vipercode{perm} feature,
which can be used to encode proof search strategies that might lead to \unsound{} inlining, as shown in \secref{sec:problem}.

\begin{figure}[t]
\begin{minipage}[t]{0.39\textwidth}
	\centering
\begin{viper2}
assume perm(x.f) $\le \frac{1}{2}$;
assert perm(x.f) $\ge \frac{1}{4}$
\end{viper2}\end{minipage}\begin{minipage}[t]{0.49\textwidth}
\centering
\begin{viper2}
exist := exist $\land$ perm(x.f) $\le \frac{1}{2}$;
if (exist) { assume perm(x.f) $\ge \frac{1}{4}$ }
\end{viper2}\end{minipage}
\figurespace{}
	\caption{The statement $s$ (\thibault{sequential composition of the two statements on the} left) is mono and framing, but rejected by the syntactic rules since it uses resource introspection. Our structural rules admit this statement. The proof obligation used to check these rules, $\mathit{guardExecs}(s, \mathit{exist})$, is shown on the right.
	}
\label{fig:instrumentation}
\end{figure}

\paragraph{Structural condition}
To validate more complex programs, we also provide \emph{structural} \gaurav{versions} of mono and framing \gaurav{that are more precise than the syntactic versions and} that can nevertheless be checked by a standard program verifier.
For simplicity, the rest of this section focuses on the structural version of mono, but the treatment of framing is analogous (see \nappref{app:structural-framing}{F}).

\peter{The structural mono property strengthens} the mono property (Def.~\ref{def:mono}) such that it can be automatically checked via a program verifier. \peter{Below, we show its definition}  for an annotation $\mathcal{A}$, a set of states $T$, and a statement $s$.
\peter{In this definition, the \emph{determinization function} $\mathit{det}$ (which maps three states and a statement to a set of states)
\thibault{corresponds to a non-empty subset of $\mathit{sem}_\mathcal{A}(\varphi_1, s)$, obtained via the process of \emph{determinization}
(explained later in this section).
In the case of a deterministic statement $s$, $\determ = \mathit{sem}_\mathcal{A}(\varphi_1, s)$,
since $\mathit{sem}_\mathcal{A}(\varphi_1, s)$ contains at most one element,
and thus it is the only subset of itself that might be non-empty.}}
\tout{is used to handle non-deterministic statements and will be explained later in this section.
For a deterministic statement $s$, $\determ$ is equal to $\mathit{sem}_\mathcal{A}(\varphi_1, s)$.}


\gaurav{
\begin{definition}\textbf{Structural mono}
\label{def:struct-mono}
	\begin{align*}
		&\mathit{structMono}_\mathcal{A}(T, s) \definedas
		\forall \varphi_1, \varphi_2 \in \Sigma \ldotp
		\varphi_1 \preceq \varphi_2 \ll T
		\land \mathit{ver}_\mathcal{A}(\varphi_1, s) \Longrightarrow \\
		&\quad \mathit{ver}_\mathcal{A}(\varphi_2, s) \land \phantom{} \tag{safety monotonicity} \\
		&\quad \left(
			\begin{array}{l}
			\forall \varphi_2' \in \mathit{sem}_\mathcal{A}(\varphi_2, s) \ldotp
			(\forall \varphi_1' \in \determ \ldotp \varphi_1' \preceq \varphi_2')
			\land \phantom{} \\
			\varnothing \subset \determ \subseteq \mathit{sem}_\mathcal{A}(\varphi_1, s)
			\end{array}
		\right) \tag{structural output mono}
	\end{align*}
\end{definition}
}
\gaurav{
The structure of the structural mono definition and the mono definition (\defref{def:mono}) are identical. The definitions differ only in the conjunct for output monotonicity.
In the original mono definition, this conjunct is given by $sem_\mathcal{A}(\varphi_1, s) \preceq sem_\mathcal{A}(\varphi_2, s)$, which, after expanding $\preceq$ \gout{\peter{according to \defref{def:order_sets}}}\gaurav{(see definition of $\preceq$ in~\secref{subsec:language})}, is equivalent to $\forall \varphi'_2 \in sem_\mathcal{A}(\varphi_2, s) \ldotp \exists \varphi'_1 \in sem_\mathcal{A}(\varphi_1, s) \ldotp \varphi'_1 \preceq \varphi'_2$. 
This formula contains an existential quantifier over states that is nested within a universal quantifier, thus making it hard to automatically reason about.\footnote{\peter{Note that the existential quantifier \thibault{hidden in $\varphi_2 \ll T$} in \defref{def:struct-mono} is not problematic for automation because it occurs on the left-hand side of an implication and is, thus, equivalent to a top-level universal quantifier.}}
}
%
%
%
%
%
%
\gaurav{The structural mono definition circumvents this forall-exists alternation issue by} \peter{strengthening the existential quantifier over $\varphi'_1$ to a universal quantifier over a non-empty range. That is, we replace the existentially-quantified formula 
$\exists \varphi'_1 \in sem_\mathcal{A}(\varphi_1, s) \ldotp \varphi'_1 \preceq \varphi'_2$ from the original definition of mono by the universally-quantified $\forall \varphi'_1 \in \determ \ldotp \varphi'_1 \preceq \varphi'_2$ (which we call the \emph{\StructInnerUniversal{}}) and the additional requirement that the range $\determ$ is \gaurav{a} non-empty subset of $\mathit{sem}_\mathcal{A}(\varphi_1, s)$.}

\gaurav{
It is easy to show that structural mono implies mono,\tout{ as we show next and} as we have \tout{also }proved in \isabelle{}:}

\thibault{
\begin{lemma}\label{lemma:struct_mono_implies_mono}\textbf{Structural mono implies mono:}
	$\mathit{structMono}_\mathcal{A}(T, s) \Longrightarrow \mathit{mono}_\mathcal{A}(T, s)$.
\end{lemma}
}

\thibault{
	\begin{proof}
		\peter{By} the similar structure between \gaurav{the definitions of} \gaurav{mono} (\defref{def:mono}) and \gaurav{structural mono} (\defref{def:struct-mono}),
		we simply need to show that (a) $\forall \varphi_2' \in \mathit{sem}_\mathcal{A}(\varphi_2, s) \ldotp
			(\forall \varphi_1' \in \determ \ldotp \varphi_1' \preceq \varphi_2')
			\land
			\varnothing \subset \determ \subseteq \mathit{sem}_\mathcal{A}(\varphi_1, s)$
		implies (b) $\mathit{sem}_\mathcal{A}(\varphi_1, s) \preceq \mathit{sem}_\mathcal{A}(\varphi_2, s)$.
		(b) is equivalent, by definition \gout{(\defref{def:order_sets})}\gaurav{(\secref{subsec:language})}, to $\forall \varphi'_2 \in sem_\mathcal{A}(\varphi_2, s) \ldotp \exists \varphi'_1 \in sem_\mathcal{A}(\varphi_1, s) \ldotp \varphi'_1 \preceq \varphi'_2$.
		We assume (a), and want to show (b).
		Let $\varphi'_2 \in sem_\mathcal{A}(\varphi_2, s)$.
		From (a), we know that $\determ$ is not empty.
		Thus, let $\varphi'_1$ be any state from $\determ$.
		Then, from (a), $\varphi'_1$ also belongs to $sem_\mathcal{A}(\varphi_1, s)$, and $\varphi'_1 \preceq \varphi'_2$ holds,
		which proves (b), and thus concludes the proof.
	\end{proof}
}

\gaurav{
The structural framing property \thibault{is} obtained \thibault{analogously} by modifying the framing definition (\defref{def:framing}). \peter{It is easy to} show that structural framing implies framing (see \nappref{app:structural-framing}{F}).
The structural condition \structcond{M}{n}{T,s} is defined identically to the semantic condition \scond{M}{n}{T,s} (\defref{def:sc}), except that the mono and framing properties are replaced by the corresponding structural properties.
We proved in \isabelle{} that the structural condition implies the semantic condition:
}

\gaurav{
\begin{theorem}
	$\structcond{M}{n}{T,s} \Longrightarrow \scond{M}{n}{T,s}$
\begin{proof}
Since \structcond{M}{n}{T,s} is defined identically to \scond{M}{n}{T,s} except that structural mono (resp.\ stuctural framing) is used instead of mono (resp.\ framing), this statement follows immediately by induction on \structcond{M}{n}{T,s} using that structural mono implies mono (Lemma~\ref{lemma:struct_mono_implies_mono}) and structural framing implies framing (Lemma~\nref{lemma:struct_framing_implies_framing}{F.2} in \nappref{app:structural-framing}{F}).
\end{proof}
\label{thm:struct_cond_implies_sem_cond}
\end{theorem}
}

%
%

\paragraph{Automating the structural condition.}
\gaurav{In the following, we explain how we check the structural condition \emph{automatically}, which boils down to automatically checking structural mono and structural framing.
Our approach is implemented by emitting additional proof obligations in \Viper{}'s verification condition generator, which builds on the \boogie{} verifier~\cite{Leino08}.
For the sake of concreteness, we will explain these additional proof obligations in terms of this implementation. 
\gaurav{However, }they can be generated in any verifier that can (a)~express an ordering on states and (b)~non-deterministically choose a state smaller than some other state.
Both requirements are met by the prevalent implementation techniques for automatic SL verifiers: verification condition generation and symbolic execution.
}
\gaurav{In our implementation, we satisfy both requirements by relying on the total-heap representation of SL states~\cite{ParkinsonSummers12} used by \Viper{}'s verification condition generator}
\gout{\thibault{In particular,} \Viper{}'s verification condition generator uses a total-heap semantics of separation logic, }where a 
state consists of a heap, a permission mask (mapping resources to the held ownership amounts), and a store of local variables. The heap and the mask are represented in \boogie{} with maps. 
This representation \gaurav{allows us to satisfy requirement} \peter{(a)~\gout{express an ordering on states}by universally quantifying over the contents of the maps representing the heaps and the masks of both states (e.g., to express that one mask contains pointwise less permission than the other), and \gaurav{requirement} (b)\gout{~non-deterministically choose a state} by picking fresh maps and then constraining them suitably via $\bcmd{assume}$ statements.
}

\peter{
\gout{With these ingredients in place}\gaurav{Given \peter{these} two requirements\pout{ expressed in the above paragraph}}, we can express the proof obligations for structural mono and framing. Both are  hyperproperties because they relate two executions of the statement $s$. As is common, we use self-composition~\cite{selfcomp} to reduce these hyperproperties to trace properties that can be checked by a standard \gout{program}verifier such as \boogie{}.}

\peter{We now explain the proof obligations for the structural mono property for a statement $s$. Checking structural framing is analogous as explained in~\nappref{app:structural-framing}{F}. We show}
\gaurav{\gout{the proof obligations for}the simpler case where $s$ is deterministic. In this case, the determinization function is given by $\determ = sem_\mathcal{A}(\varphi_1, s)$. At the end of this section, we will then explain how to handle \peter{non-deterministic statements}.
}

%

\begin{figure}[t]
	\centering
	\begin{algorithmic}[1]
		\footnotesize
		\IF{(*)} \label{line:if}
			\STATE Let $\varphi_1$, $\varphi_2$ be \Viper{} states s.t.\ 
			$\varphi_1 \preceq \varphi_2 \preceq \mathit{currentViperState}$
			\label{line:states}
			\STATE $(\mathit{exist}, \mathit{currentViperState})  \leftarrow (\top, \varphi_1)$ \label{line:start1}
			\STATE $\mathit{guardExecs}(s, \mathit{exist})$  \label{line:middle1}
			\STATE $\varphi'_1 \leftarrow \mathit{currentViperState}$ \label{line:end1}
			\STATE $\mathit{currentViperState} \leftarrow \varphi_2$ \label{line:start2}
			\STATE $s$ \label{line:middle2}
			\STATE \textbf{assert} $\mathit{exist} \land \thibault{\varphi'_1 \preceq \mathit{currentViperState}}$ \label{line:check}
			\STATE \textbf{assume} \textit{false} \label{line:stop}
		\ENDIF
\end{algorithmic}
	\caption{Proof obligation, expressed via self-composition, to \gaurav{automatically} check if a deterministic statement $s$ is structurally mono. \peter{We use pseudocode here, but the check can be expressed directly in \Viper{}'s verification condition generator based on \boogie{}.}}
	\label{fig:algo}
\end{figure}

\peter{Our structural condition requires the call-free statements between method calls to be structurally mono. To check this property, we precede each such statement $s$ with the code shown in \figref{fig:algo}, which generates the necessary proof obligations.}
Note that \peter{this code is included} in a non-deterministic branch (line~\ref{line:if}), which is killed after the check (line~\ref{line:stop}). This allows us to include the check in the encoding of the inlined program without affecting \peter{the rest of its verification}.

\peter{According to \defref{def:struct-mono}, structural mono is defined relative to an upper bound $T$, which \gaurav{in the structural condition is instantiated with} the set of states reachable before the statement $s$ in the inlined program. These states are implicitly represented by the current verification state of the \Viper{} program before statement $s$; in \figref{fig:algo}, we refer to this state as $\mathit{currentViperState}$; like all our states, it consists of a heap, a permission mask, and a store.}

\peter{To prove structural mono for all states $\varphi_1, \varphi_2$, we choose (in line~\ref{line:states}) fresh states non-deterministically and constrain them as prescribed by \defref{def:struct-mono} (where the current verification state represents $\varphi_3 \in T$). \Viper{} does not have a built-in order on states but, as we explained above, we can express this easily via quantification over the contents of the heap and the mask.} 

\peter{Structural mono may assume that $s$ verifies successfully in state $\varphi_1$ ($\mathit{ver}_\mathcal{A}(\varphi_1, s)$ in \defref{def:struct-mono}).
We achieve this by setting the current state to $\varphi_1$ (line~\ref{line:start1}), execute $s$ (line~\ref{line:middle1}), and record the final state as $\varphi'_1$ (line~\ref{line:end1}).
However, since the successful verification of $s$ in $\varphi_1$ is an \emph{assumption} of structural mono, we need to catch situations where this verification fails.
In that case, we would incorrectly report an error even though structural mono is not violated.
We also need to detect if the execution of $s$ is infeasible because this would make the remaining proof obligations hold vacuously\thibault{, even though
structural mono is actually violated in this case.}}
We solve both issues by executing a modified version \thibault{of $s$, namely $s' \triangleq \mathit{guardExecs}(s, \mathit{exist})$,}
that avoids \emph{infeasible} executions by accumulating assumptions using a fresh boolean variable $\mathit{exist}$, and by executing statements in $s$ only if $\mathit{exist}$ holds. $s'$ avoids \emph{failing} executions by turning assertions into assumptions. \gaurav{Hence}, after the execution of $s'$, if $\mathit{exist}$ holds, then $\varphi_1'$ corresponds to an output state of a verifying execution of $s$ in $\varphi_1$, \peter{as required by \defref{def:struct-mono}.}
We illustrate the transformation $\mathit{guardExecs}(\_, \mathit{exist})$ on the right of \figref{fig:instrumentation}.

After this first (instrumented) execution of $s$ in the state $\varphi_1$, we execute the (non-instrumented) statement $s$ in the state $\varphi_2$ (lines~\ref{line:start2} to~\ref{line:middle2}).
If no error is reported during this execution, we can guarantee that $s$ is \emph{safety} monotonic for the bound $T$.
\peter{We check whether $s$ is also \emph{\thibault{structural} output monotonic}\thibault{,
which corresponds to the second conjunct on the implication's right-hand side of the structural mono definition (marked by "structural output mono." in \defref{def:struct-mono})}, as follows.
Since we assume $s$ to be deterministic (and, thus, $\determ = sem_\mathcal{A}(\varphi_1, s)$),
\defref{def:struct-mono} requires \thibault{$sem_\mathcal{A}(\varphi_1, s) \neq \varnothing$.
In other words, it requires}
the first execution of $s$ to reach a final state, that is, to be feasible, which we check by asserting that $\mathit{exist}$ holds (line~\ref{line:check}).
Moreover, the final state of the second execution of $s$ ($\varphi_2'$ in \defref{def:struct-mono}) must be larger than the final state of the first ($\varphi_1'$ in \defref{def:struct-mono}), which we assert as well.}
Note that the \gaurav{\StructInnerUniversal} must hold only for $\varphi'_2 \in sem_\mathcal{A}(\varphi_2, s)$. \peter{This is automatically the case in our proof obligation:}
if the execution of $s$ on line~7 is infeasible, then the check on line~8 holds trivially.

\peter{In summary, the proof obligation from \figref{fig:instrumentation} reflects directly the definition of structural mono for deterministic statements $s$. Even though \defref{def:struct-mono} expresses a non-trivial hyperproperty that compares entire states, the resulting proof obligations can be proved automatically using standard verification tools. \gout{It remains to discuss how to check the property for non-deterministic statements, as we do next.}\gaurav{Next, we discuss how to check the property for non-deterministic statements.} }

\paragraph{Determinization.}
\thibault{The determinization function	
$\determ$ used in \defref{def:struct-mono}
yields the subset of final states $\mathit{sem}_\mathcal{A}(\varphi_1, s)$ of executions that make, wherever possible, the same non-deterministic choices as the execution that starts in $\varphi_2$ and ends in $\varphi'_2$. Here, the states correspond to the states from \defref{def:struct-mono}; in particular, $\determ$ is meaningful when $\varphi_1 \preceq \varphi_2$. 
The corresponding proof obligation when $s$ is deterministic (and thus
$\determ = sem_\mathcal{A}(\varphi_1, s)$)
is shown in \figref{fig:algo}.
}

\thibault{However, for non-deterministic statements $s$, using the entire set of final states $\mathit{sem}_\mathcal{A}(\varphi_1, s)$ would lead to an overly strong definition of structural mono: \defref{def:struct-mono} would compare final states $\varphi'_1$ and $\varphi'_2$
obtained by making different non-deterministic choices and, for that reason, fail to satisfy $\varphi'_1 \preceq \varphi'_2$. 
To avoid this problem, determinization aligns the two executions of $s$, to obtain 
a more relevant \thibault{(and smaller)} subset $\determ$
of $sem_\mathcal{A}(\varphi_1, s)$
(recall that the \gaurav{\StructInnerUniversal} is $\forall \varphi'_1 \in \determ \ldotp \varphi'_1 \preceq \varphi'_2$).}
Instead of comparing every pair of executions $(E_1, E_2)$ \peter{of statement $s$} starting in $\varphi_1$ and $\varphi_2$\thibault{ and ending in $\varphi_1'$ and $\varphi_2'$}, respectively,
determinization compares \peter{only those executions that resolve non-determinism similarly. To achieve this, we record} all non-deterministic choices (such as the initial values of variables and newly-owned heap locations, or the existential fractional ownerships that have been chosen)
made during the execution $E_1$ (line~4 in \figref{fig:algo}), and then ignore the pair $(E_1, E_2)$ if $E_2$ (line~7) resolves non-determinism in a different way,
provided that there is (at least) one pair with $E_2$ that is not ignored.
\peter{The latter proviso} ensures that the set of executions $E_1$ that are compared with $E_2$ is non-empty\thibault{, \ie{} that $\determ \neq \varnothing$, as required by the structural mono property (\defref{def:struct-mono})}.
See \nappref{app:determinization}{G} for more details.

\section{Evaluation}\label{sec:evaluation}
In this section, we evaluate four important aspects of our technique. We demonstrate that
(1)~features that may cause inlining to be \unsound{} are widely used,
(2)~\unsound{} inlining actually occurs in practice,
(3)~our structural condition is sufficiently precise, that is, it captures most examples that violate the syntactic condition but can be inlined in a \longsound{} manner, and
(4)~our implementation of \sound{} inlining in \Viper{} effectively detects bugs.

Our evaluation considers the test suites from \verifast{} (1002 files), \grasshopper{} (314 files), \nagini{} (232 files), and \rslviper{} (14 files); the latter two encode verification problems into \Viper{}. All four verifiers are interesting subjects for our evaluation because they use automation techniques and generate proof obligations for features that potentially lead to \unsound{} inlining and that violate our syntactic condition. Examples from other tools, such as Prusti~\cite{AstrauskasMuellerPoliSummers19b}, can always be validated by our syntactic condition, which demonstrates the usefulness of this check, but makes those tools less relevant for our evaluation.

We have implemented \longsound{} inlining for loops and method calls in \Viper{}, taking partial annotations into account.
Our implementation automatically checks the structural  condition using the technique described in~\secref{sec:automation} (and omits it when the syntactic condition holds).
Both inlining and checking the structural condition are performed by extending \Viper{}'s verification condition generator, which translates \Viper{} programs to \boogie{}~\cite{Leino08}.
The tool and examples \thibault{are part of our publicly available artifact~\cite{artifact}}.

In the following, we refer to files (resp.\ features) that violate the syntactic condition as \emph{non-trivial files} (resp.\ \emph{non-trivial features}).

\subsection{The Syntactic Condition is Often Violated}

\renewcommand\DTstyle{\rmfamily \small}

\begin{table}[t]
  \caption{
    \thibault{Results of our syntactic checks and subsequent manual analysis for files from the test suites of \rslviper{} (R), \nagini{} (N), \grasshopper{} (G), and \verifast{} (VF).
  The results show that 67\% of the tests include features that our syntactic checks do not capture,
    that inlining may be \unsound{} for the methods in 31\% of the \gaurav{manually} analyzed test cases, and that our structural condition is sufficiently precise to validate 94\% of the test cases that are \alwayspreserving{}.}
  }
  \small
  \centering
  \begin{tabular}{l||c|c|c|c|c|c}
    & ~~\textbf{R}~~ & ~~\textbf{N}~~ & ~~\textbf{G}~~ & ~~\textbf{VF}~~ & \textbf{Total} \\
    \hline
    \multirow{6}{*}{%
    \resizebox{0.44\textwidth}{!}{
    \begin{minipage}{7.1cm}\dirtree{%
      .1 All files.
      .2 Satisfy syntactic condition.
      .2 Violate syntactic condition.
      .3 Manually analyzed.
      .4 Not \alwayspreserving{}.
      .4 \Alwayspreserving{}.
      .5 Validated by semantic condition.
      .5 Validated by structural condition.
      .2 Lines of code (mean / median).
      }\end{minipage}
    }}
    & 14 & 232 & 314 & 1002 & 1562 \\
    & - & - & 203 & \thibault{306} & \thibault{509} \\
    & 14 & 232 & 111 & \thibault{696} & \thibault{1053} \\
    & 12 & 20 & 20 & 20 & 72 \\
    & 8 & 4 & 2 & 8 & 22 \\
    & 4 & 16 & 18 & 12 & 50 \\
    & 4 & 15 & 18 & 12 & 49 \\
    & 2 & 15 & 18 & 12 & 47 \\
    & \gaurav{85 / 104} & \gaurav{73 / 47.5} & \gaurav{124 / 57} & \gaurav{160 / 67} & \gaurav{139 / 61}
  \end{tabular} 
  \label{tbl:results}
\end{table}

\gaurav{The first three rows in } \tabref{tbl:results} \gaurav{show} that non-trivial features appear often in the analyzed test suites: Out of 1562 files in total, \thibault{1053} (67\%) violate the syntactic condition
(\peter{the numbers were} obtained via automatic detection of non-trivial patterns\pout{; the detection typically takes less than a second}).
This shows not only that most files contain features that might make inlining \unsound, but also that our structural condition, which is more fine-grained than the syntactic one, is indeed necessary to determine whether inlining is \sound.
Proof rule selection strategies that depend on the owned resources are applied in 5 \rslviper{} files (out of 14, 36\%) and 64 \nagini{} files (out of 232, 28\%).
Moreover, we found 111 non-trivial files in \grasshopper{} (out of 314, 35\%) and \thibault{696} in \verifast{} (out of 1002, \gaurav{69}\%), mostly because of imprecise assertions that appear in predicate bodies, specifications, and ghost code.
Apart from these two scenarios that we have discussed throughout the paper, there is a third scenario that violates the syntactic condition.
11 \grasshopper{} files (out of 314, 4\%) and all \nagini{} files contain assertions that specify exact bounds on the resources owned. 
\grasshopper{}'s \vipercode{assert R} statement succeeds iff the method owns \emph{exactly} the resources specified by \code{R}, reflecting \grasshopper{}'s underlying classical SL\@.
\nagini{} asserts at the end of each method that there are no remaining obligation resources to release a lock, which would get leaked when the method terminates.
Since inlining affects the resources owned, these instances can also lead to \unsound{} inlining.


\subsection{\UnSound{} Inlining Occurs in Practice}\label{subsec:unsound_inlining_in_practice}


Examples that violate our syntactic condition do not necessarily make inlining \unsound. 
To assess whether inlining is actually \unsound{} for non-trivial examples, we further analyzed non-trivial files from the four verifiers.
\gaurav{Since most \peter{methods} in the test suites have no (or very few) clients that invoke \peter{them}, it would be insufficient to check whether \unsound{} inlining occurs only for \peter{the} existing clients as the initial statement, or for a fixed selection of inlining bounds.
Instead, we \peter{(manually) analyze the methods for any possible} client code (with minor restrictions) and any inlining bound.}
\gout{We classified}\gaurav{In particular, we classify} a file as \emph{\alwayspreserving{}} if inlining every method in every caller context that satisfies the syntactic condition (that is, does not itself make inlining \unsound)\footnote{For \nagini{}, we used 
further restrictions on the clients to avoid that systematic leak checks prevent \emph{all} examples from being \alwayspreserving{}, which would not faithfully reflect typical clients.}
is \longsound{} for every inlining bound.
In \rslviper{} and \nagini{}, we analyzed the methods in the corresponding \Viper{} encoding.
For 114 files in \nagini{}, we could automatically deduce that they were \alwayspreserving{} using extended verifier-specific syntactic checks.
For the remaining files, manual inspection was required.
For all verifiers except \grasshopper{}, some of these files were too complex for manual inspection, so we automatically discarded all those files with too many complicated features.
This still left us with a large and diverse set of examples (12 for \rslviper{}, 79 for \nagini{}, 111 for \grasshopper{}, \thibault{271} for \verifast{}).
From these examples, we picked 20 examples randomly for each verifier (except \rslviper{}, where we took all).
The results of this manual analysis are presented in \tabref{tbl:results}.
We took existing annotations into account. Not doing so would have resulted in a different classification for only 2 examples in \verifast{}, which would have been classified as \alwayspreserving{} instead.

Overall, out of the 72 non-trivial files that we analyzed manually, 22 \tout{examples}\thibault{files} (31\%) are not \alwayspreserving{}.
This shows that inlining is \tout{indeed }\unsound{} in a non-negligible number of cases, and that inlining is \sound{} in the majority of cases even when the syntactic condition is violated; thus, our more-precise structural condition is needed to validate those (see \secref{sec:structural_effective}).

Our manual inspection revealed that in \verifast{}, \unsound{} inlining is often due to imprecise assertions.
In \rslviper{}, the \unsound{} pattern from~\figref{fig:rsl-intro-unsound} occurs in 5 examples.
In \grasshopper{} and \nagini{}, the main source of \unsound{} inlining are assertions on exact bounds of resources, which are \unsound{} in calling contexts that provide more resources\pout{ than expected by the exact resource bound}.

\gout{
(paragraph moved to ``threats to validity'')
We took the following measures to increase the confidence in our manual classification.
In cases where a file is not always preserving, we wrote a simple client satisfying the syntactic condition that invokes a method in the file to confirm this observation.
That is, we (1)~wrote an annotation for which the program (consisting of the client as the initial statement) verifies modularly and (2)~identified an inlining bound for which the inlined program does not verify.
To check (2), we used our inlining tool for \Viper{}-based verifiers; for \grasshopper{} and \verifast{}, we inlined calls and unrolled loops manually, and then used the corresponding verifier.
In cases where inlining is always preserving for a file, we sketched informal proofs for every method in the file.
We did so by considering each statement of a method that does not satisfy the syntactic condition, and reasoned why (given the rest of the method) it cannot lead to non-preserving inlining.
}

\subsection{Our Conditions are Effective and Precise}
\label{sec:structural_effective}

\begin{table}[t]
  \caption{Non-trivial examples from the test suites of \nagini{} (N), \rslviper{} (R), \verifast{} (VF), \grasshopper{} (G). \verifast{} and \grasshopper{} examples were translated manually to \Viper{}.
We show the lines of code and lines of annotations needed for successful modular verification.
The next three columns indicate whether inlining is \sound{} (Inl.P.), the semantic condition holds (SC), and the structural condition holds (Str.C.).
The verification time with inlining (T) is the average of 5 runs on a Lenovo T480 with 32 GB, i7-8550U 1.8 GhZ, Windows 10.
The last two columns show the number of seeded errors (\#Err.) to be found with inlining and the number of spurious errors reported when verifying modularly, but without annotations (\#Sp.Err.).
\gaurav{If more than one initial statement was considered in which calls were inlined, then the number of spurious errors is given by the average of spurious errors reported for each of the initial statements.}
%
  }
  \figurespace{}
  \small
  \centering
  \resizebox{0.9\textwidth}{!}{
  \begin{tabular}{l|c|c|c|c|c|c|c|c|c}
  Name & LOC & Ann. & Inl.P. & SC & Str.C. & T [sec] & \#Err. & \#Sp.Err.\\
  \hhline{=|=|=|=|=|=|=|=|=|=}
  $N_1$: iap\_bst & 122 & 22 & \tick & \tick & \tick & 19.2 & 2 & 10.3   \\
  $N_2$: parkinson\_recell & 37 & 9 & \tick & \tick & \tick & 11.9 & 3 & 5.4  \\
  $N_3$: watchdog & 52 & 9 & \tick & \tick & \tick & 10.8& 1 & 3  \\
  $N_4$: loops\_and\_release & 20 & n/a & \xmark & \xmark & \xmark & 9.0 & n/a & n/a  \\
  $R_1$: rust\_arc~\cite{DokoVafeiadis17} & 26 & 6 & \tick & \tick & \tick & 3.4 & 7 & 1.6  \\
  $R_2$: lock\_no\_spin & 17 & 2 & \tick & \tick & \xmark & 42& 0 & 1  \\
  $R_3$: msg\_pass\_split\_1 & 10 & 3 & \tick & \tick & \tick & 2.7& 1 & 5  \\
  $R_4$: msg\_pass\_split\_2 & 10 & n/a & \xmark & \xmark & \xmark & 5.8& n/a & n/a  \\
  $\mathit{VF}_1$: account & 43 & 8 & \tick & \tick & \tick & 2.2 & 2 & 3.7  \\
  $\mathit{VF}_2$: lcp~\cite{lcp_vf} & 54 & n/a & \xmark & \xmark & \xmark & 7.3 & n/a & n/a  \\
  \gaurav{$\mathit{VF}_3$:} \gaurav{iterator} & \gaurav{49} & \gaurav{8} & \tick & \tick & \tick & \gaurav{1.6} & \gaurav{2} & \gaurav{4.5} \\
  \gaurav{$\mathit{VF}_4$:} \gaurav{stack} & \gaurav{50} & \gaurav{6} & \tick & \tick & \tick & \gaurav{1.9} & \gaurav{3} & \gaurav{2.5} \\
  $G_1$: bstree & 100 & n/a & \xmark & \xmark & \xmark & 13.5 & n/a & n/a \\
  \gaurav{$G_2$:} \gaurav{nodes} & \gaurav{54} & \gaurav{29} & \tick & \tick & \tick & \gaurav{1.4} & \gaurav{3} & \gaurav{7.8}
  \end{tabular}}
  \label{tbl:examples}
\end{table}

Our semantic and structural conditions are sufficient for inlining to be \sound, but not necessary.
To evaluate the precision of the conditions, we further \gout{analyzed}\gaurav{examined} \tout{the \gaurav{analyzed} }examples
for which inlining is \alwayspreserving{}.
For the 114 automatically handled examples in \nagini{}, we could also automatically deduce that both the semantic and structural conditions hold.
For the \gaurav{72 manually handled} examples\gout{in~\tabref{tbl:results}}, we \gaurav{also} evaluated the conditions manually and will discuss the results \gaurav{(shown in~\tabref{tbl:results})} next.
We also assessed the usefulness of the bounded relaxation of our condition.

\smallParagraph{Our semantic condition is \tout{sufficiently }precise.}
As shown by the results of our \thibault{manual} analysis (\tabref{tbl:results}),
our semantic condition captures almost all (49 out of 50, 98\%) non-trivial files classified as \alwayspreserving{}.
Besides the \alwayspreserving{} file (in \nagini{}) not captured by our condition,
there are methods in the \unsound{} \tout{test cases}\thibault{files}
for which inlining is \unsound{} in \emph{some} caller contexts, but \sound{} in others.
Our manual inspection revealed some \sound{} caller contexts that our semantic condition cannot validate.
We found such patterns in \rslviper{} (see \figref{fig:rsl-intro-unsound}), \verifast{}, and \nagini{}, but not in \grasshopper{}.
Dealing with these patterns requires a non-compositional approach taking the entire program into account, which is practically infeasible.


\smallParagraph{Our structural condition is effective.}
Our structural condition \gout{approximates}\gaurav{is stronger than} the semantic condition. 
\tabref{tbl:results} shows that this \peter{over-}approximation is very precise in practice: the structural condition validates 96\% of the \alwayspreserving{} files that
satisfy the semantic condition.

\smallParagraph{The bounded relaxation is required.}
We argued in \secref{sec:overview-resource-bound} that the bounded relaxation of our condition is needed to validate common patterns.
Our manual inspection confirmed this claim. For example, \nagini{} uses two patterns (occurring in 62 and all 232 test cases, resp.) that can be validated only with the bounded conditions.
In \grasshopper{} and \verifast{}, imprecise assertions (occurring in 107 and all 697 test cases, resp.) are often unambiguous in the context they are inlined in,
and thus satisfy our structural condition only because of the bounded relaxation.

\gout{
(paragraph moved to ``threats to validity'')
\smallParagraph{Impact of automatically discarding files.}
As we discussed in~\secref{subsec:unsound_inlining_in_practice}, we discarded files automatically before choosing files randomly for manual inspection.
Nevertheless, we believe that our manual analysis in~\tabref{tbl:results} (and accompanying automatic analysis for \nagini{}) reflects the big picture of the test suites.
In \verifast{}, we discarded files that were too large for manual inspection. 
Features that do not satisfy the syntactic condition in \verifast{} often appear within small recurring patterns that also show up in the considered smaller files.
Therefore, we are confident that our results for ``\alwayspreserving{}'' files would be similar for the discarded files.
Since the discarded files are larger, we likely would get more files that exhibit \unsound{} inlining, which would still support our conclusion that \unsound{} inlining occurs.
For \nagini{} and \rslviper{}, we discarded files because they contained features that were too complex to analyze manually.
Thus, we cannot argue that our results from~\tabref{tbl:results} translate.
However, in both cases, we still consider the vast majority of the test cases (83\% and 86\%, respectively), which means our results reflect a majority of the test suites.
}


\subsection{\Sound{} Inlining Effectively Detects Bugs}
\label{sec:usefulness_inlining}

\smallParagraph{The structural condition can be checked automatically.}
The previous subsection showed that our structural condition is sufficiently precise.
To evaluate whether it can be checked automatically, we manually selected, out of the 1053 non-trivial files, a diverse set of examples (shown in~\tabref{tbl:examples}) that reflect the non-trivial patterns occurring in the different verifier test suites and that could be translated to \Viper{}. Our tool correctly reports whether the structural condition holds in all cases irrespective of whether existing annotations are taken into account.

\smallParagraph{Inlining detects errors effectively.}
To evaluate how effective inlining is in finding true errors without annotations,
we consider all examples in \tabref{tbl:examples} for which inlining is \sound, and some examples taken from the \Viper{} test suite, most of which satisfy the syntactic condition (\ntabref{tbl:syntactic_examples}{3} in \nappref{app:syntactic_evaluation}{H}).
We seeded errors by \tout{either }making simple changes in the implementations or writing clients that use methods
incorrectly.
\gaurav{For several examples, we considered more than one initial statement in which calls were inlined and loops were unrolled.}
Our tool was able to report every true error for some sufficiently large inlining bound \gaurav{(bounds between 1 and 4 were sufficient for all examples except for $N_4$ and $G_2$ in~\tabref{tbl:examples}, where bounds 10 and 11 were required since both examples contain a loop that iterates 10 times)}. 
Our tool never reported a spurious error, which was expected since our structural condition is sufficient to ensure that inlining is \longsound{}.

\smallParagraph{Inlining reduces annotation overhead.}
Verifying the same examples modularly without providing any annotation results in at least one spurious error each, and 3 on average.
To assess the amount of annotations saved by using inlining instead of modular verification, we considered annotations for all examples to successfully verify without inlining (that is, to show they are memory safe and satisfy all provided assertions). This required \gout{213}\gaurav{256} lines of annotation in the programs for \gout{999}\gaurav{1152} lines of code, which inlining does not require.
This result shows that inlining is useful to find true errors with low annotation overhead and to gain confidence that an implementation is correct.

\gaurav{
\subsection{Threats to Validity}
We identified the following threats to the validity of our evaluation.
}

\gaurav{
\smallParagraph{\thibault{Dataset}.}
}
\gaurav{\peter{Our examples, which we took from} the test suites of \verifast{}, \grasshopper{}, \nagini{}, and \rslviper{}, \peter{might not be representative of realistic code.}
We believe this threat to \peter{be minor} since (1)~these test suites contain \peter{various practically-relevant verification problems} and (2)~\peter{contain} non-trivial specifications and exercise features that show up in real-world programs.
}

As we discussed in~\secref{subsec:unsound_inlining_in_practice}, we discarded \peter{some} files automatically before choosing files randomly for \gout{manual inspection}\gaurav{the manual analysis presented in~\secref{subsec:unsound_inlining_in_practice} and~\ref{sec:structural_effective}}.
\peter{We are convinced that discarding these files does not compromise the validity of} our manual analysis\gout{in~\tabref{tbl:results}} (and accompanying automatic analysis for \nagini{}) \peter{for the following reasons}.
In \verifast{}, we discarded files that were too large for manual inspection. 
Features that
\thibault{violate} the syntactic condition in \verifast{} often appear within small recurring patterns that also show up in the considered smaller files.
We are \thibault{thus} confident that our results for ``\alwayspreserving{}'' files would be similar for the discarded files.
Since the discarded files are larger, we likely would get more files that exhibit \unsound{} inlining, which would still support our conclusion that \unsound{} inlining occurs.
For \nagini{} and \rslviper{}, we discarded files
\tout{because they}%
\thibault{that contain} features that were too complex to analyze manually.
\gout{Thus, we cannot argue that our results from~\tabref{tbl:results} translate.}
However, in both cases, we still consider the vast majority of the test cases (83\% and 86\%, respectively)\pout{, which means our results reflect a majority of the test suites}.

\gaurav{
\smallParagraph{\tout{Correctness of m}\thibault{M}anual analysis.}
}
\peter{We may have made mistakes  in our manual analysis presented in~\secref{subsec:unsound_inlining_in_practice} and~\ref{sec:structural_effective}. We mitigated this risk as follows.}
\gaurav{When} a file is not \alwayspreserving{}, we wrote a simple client satisfying the syntactic condition that invokes a method in the file to confirm this observation.
That is, we (1)~wrote an annotation for which the program (consisting of the client as the initial statement) verifies modularly and (2)~identified an inlining bound for which the inlined program does not verify.
To check (2), we used our inlining tool for \Viper{}-based verifiers; for \grasshopper{} and \verifast{}, we inlined calls and unrolled loops manually, and then used the corresponding verifier.
\gout{In cases where}\gaurav{When} inlining is \alwayspreserving{} for a file, we sketched informal proofs for every method in the file.
We did so by considering each statement of a method that does not satisfy the syntactic condition, and reasoned why (given the rest of the method) it cannot lead to non-preserving inlining.
\gaurav{We sketched similar proofs in cases where the semantic condition and structural condition hold.}

\gaurav{
\smallParagraph{Error seeding.}
We seeded the errors in~\tabref{tbl:examples} and \ntabref{tbl:syntactic_examples}{3} in \nappref{app:syntactic_evaluation}{H} mostly ourselves. We tried to mitigate a potential bias \tout{in the errors }by seeding different kinds of errors 
\gaurav{(\eg asserting incorrect properties in clients, using libraries incorrectly, erroneously adjusting library implementations)}.
}

\section{Related Work}\label{sec:related_work}
\corral{}~\cite{LalQL12} detects bugs in C programs by translating them to \boogie{}~\cite{Leino08}. The \tout{deductive }\boogie{} verifier is used to check correctness of the inlined \boogie{} program.
Inlining is trivially \longsound{} for \corral{}.
\gout{In addition to inlining}\gaurav{Additionally}, \corral{} applies various techniques to improve efficiency such as approximating method calls with inferred \emph{method call summaries} without inlining them.
\citet{LourencoFradePinto19} consider \gout{modular verification and}bounded verification using inlining in the context of a standard verification language without any resources.
They directly connect correctness of the original program (instead of verification w.r.t.\ a verifier semantics) to verification of the inlined program, which is not possible in our setting due to \gout{the potentially incomplete}proof search algorithms.

\gout{One motivation for our work}\gaurav{One of our motivations} as mentioned in~\secref{sec:introduction} is to use bounded verification as a stepping stone for modular verification.
This is also the case for \citet{BeckertJMLBMC20}, who define a translation from a Java program with annotations expressed in the Java Modeling Language\gout{(JML)}~\cite{LeavensJML2006} to a Java program that is accepted by the \textsc{JBMC} bounded model checker~\cite{CordeiroJBMC2018}.
Bounded model checkers (BMC) such as \textsc{JBMC} and \textsc{CBMC}~\cite{ClarkeKroeningLerda04} perform bounded verification (via inlining) and detect errors effectively but do not support annotations such as method contracts and frame conditions, and generally support less expressive assertions than deductive verifiers.

Instead of using an off-the-shelf BMC, we inline the program and then use already-existing automatic deductive SL verifiers, which have a mature automation infrastructure for SL assertions.
This allows us to directly support inlining partial annotations or calls to library methods without additional work on SL assertion support in BMC\@. 
Moreover, performing both kinds of verification within the same tool ensures that no verification errors are caused by switching from one tool to the other, for instance, due to small differences in the verifier semantics.
Nevertheless, it would be interesting to explore a BMC technique that supports the handling of SL assertion logics such as those from \grasshopper{}, \Viper{} and \verifast{} (potentially building on SL runtime checking~\cite{NguyenKC08,Agten0P15} or SL model checking~\cite{BrotherstonGKR16}).

\thibault{While safety monotonicity and the framing property have been studied in the context of \gaurav{SL}~\cite{Yang2002,Calcagno2007},
the relaxed conditions we use (where the states these conditions quantify over are bounded using the inlined program's execution)
have, to the best of our knowledge, not been explored before.}
\tout{Both safety monotonicity and the framing property have been studied in the context of \gout{separation logics}\gaurav{SL}~\cite{Yang2002,Calcagno2007}.
\gout{In our work,}\gaurav{We} use the inlined program's execution to bound the states that these conditions quantify over, which results in relaxed conditions.
To \gout{the best of}our knowledge, such a relaxation has not been explored\gout{before}.}
As shown in~\secref{sec:evaluation}, these relaxations are essential to capture common idioms.
\tout{Moreover, our \soundness{} condition is more than just a bounded version of
the \emph{local action} property in SL\tout{~\cite{Calcagno2007}}.}
\thibault{Moreover,
our \soundness{} condition imposes} monotonicity and framing constraints on \emph{different} parts of the program
(based on the relationship between modular and inlined verification),
which is
\tout{This difference is }crucial to capture common patterns that satisfy these properties
but contain statements that do \emph{not}, as illustrated in \secref{sec:overview}\gout{and shown in \secref{sec:evaluation}}.
Finally, we propose a novel output monotonicity property, which, to the best of our knowledge, has not been used in the 
context of SL\@.

Several automatic SL verifiers use incomplete heuristics in their proof search strategies that may lead to \unsound{} inlining,
such as \caper{}~\cite{caper17}, \grasshopper{}~\cite{grasshopper}, \nagini{}~\cite{EilersMueller18}, \refinedc{}~\cite{Sammler21}, \rslviper{}~\cite{SummersMueller18}, \steel{}~\cite{FromherzRSGMMR21}, \vercors{}~\cite{BlomHuisman14}, \verifast{}~\cite{Jacobs-Verifast11}, and \Viper{}~\cite{MuellerSchwerhoffSummers16}.
\thibault{\refinedc{} uses incomplete rules \gaurav{and} \steel{} uses incomplete heuristics
to instantiate existentially-quantified variables;
both may lead to \unsound{} inlining (\nfigref{fig:refinedc}{13} and \nref{fig:steel}{14} in \nappref{app:examples}{I}).}
\caper{} uses backtracking when resolving non-deterministic choices to make the proof search more complete.
However, for the \emph{region creation} proof rule, which can (but need not) be applied at various points, \caper{} cannot explore all options.
Instead, it uses incomplete heuristics that can lead to \unsound{} inlining \thibault{(\nfigref{fig:caper}{12} in \nappref{app:examples}{I})}.
While \gout{the framework we present here}\gaurav{our framework}
can be applied to
\grasshopper{}, \verifast{}, and verifiers based on
\Viper{},
it cannot be applied to \caper{}, because backtracking does not fit into our formal model.
\section{Conclusion}\label{sec:conclusion}
We demonstrated that inlining may introduce false positives when using automatic SL verifiers.
Their automation techniques are sensitive to changes in ownership, which occur inevitably during inlining.
We identified a novel compositional \soundness{} condition and proved that it is sufficient to ensure \longsound{} inlining.
Since this condition is difficult to check, we developed two approximations that can be checked syntactically and with a standard program verifier, respectively.
Our evaluation shows that these conditions are necessary and capture most use cases.

Our work paves the way to bounded verification within automatic SL verifiers without the risk of false positives. 
We believe that the foundations presented in this paper can also be used for other applications, such as the caching of verification results in automatic SL verifiers. 
Existing caching approaches~\cite{LeinoW15} do not re-verify code after a call if the postcondition of the callee is strengthened.
Such techniques may be unsound when applied to automatic SL verifiers if the statement after the call is not safety monotonic.
One direction for future work is to devise a sound caching approach for automatic SL verifiers using our techniques.

\begin{acks}
We thank the following people for their help:
Vytautas Astrauskas and Marco Eilers (\nagini{}),
Felix Wolf (\caper{}),
Michael Sammler (\refinedc{}),
Denis Merigoux (\steel{}),
Malte Schwerhoff (\verifast{}),
and
Christoph Matheja (giving feedback on previous drafts).
This work was partially funded by the Swiss National Science Foundation (SNSF) under
Grant No. 197065.
\end{acks}


\section*{Data Availability Statement}
Our publicly-available artifact~\cite{artifact} contains:
\begin{enumerate}
    \item \isabelle{} proofs of the technical results from \secref{sec:soundness} and~\ref{sec:automation}.
    \item An analysis of the test suites of \grasshopper{}, \nagini{}, \rslviper{}, and \verifast{}, corresponding to the results shown in \tabref{tbl:results}.
    \item The inlining tool for \Viper{}, described in \secref{sec:evaluation}, which inlines calls and unrolls loops, while also checking the structural condition.
    \item A test framework that runs the inlining tool on the examples from \tabref{tbl:examples} and \ntabref{tbl:syntactic_examples}{3}.
\end{enumerate}
%

\clearpage
\bibliography{references}

\clearpage


%
\appendix
\section{Separation Algebra, Parametric Language, and Semantics}~\label{app:soundness}

\subsection{States as Elements of a Separation Algebra with a Store}

To precisely capture the semantics of loops and method calls in verifiers such as \grasshopper{}, \verifast{}, or \Viper{}, our state model builds on a separation algebra~\cite{Calcagno2007,Dockins2009} to represent resources.
In this separation algebra, local variables are represented as pure (infinitely duplicable) resources.
The store of local variables (that is, the pure resources of a state) is referred to as the \emph{core} of a state, with a similar notation as in \textsc{Iris}~\cite{Jung2018}.
All resources that are not local variables, such as heap locations, are non-duplicable and referred to as \emph{impure} resources.

\begin{definition}
	A \textbf{partial commutative monoid} is a triple $(\Sigma, \oplus, u)$
	where
	$\Sigma$ is a set of states,
	$\oplus: \Sigma \times \Sigma \rightarrow \Sigma$ is a partial operation that is commutative and associative,
	and $u \in \Sigma$ is a neutral element for $\oplus$.
	The \textbf{induced partial order} $\preceq$ on elements of $\;\Sigma$ is defined as
	$\varphi_1 \preceq \varphi_2 \Longleftrightarrow (\exists c \in \Sigma \ldotp \varphi_2 = \varphi_1 \oplus c)$,
	and we write $\varphi_1 \# \varphi_2$ if and only if $\varphi_1 \oplus \varphi_2$ is \textbf{defined}.
	A state $\varphi$ is \textbf{pure} if and only if $\varphi = \varphi \oplus \varphi$.
	The \textbf{core} of a state $\varphi$, written $|\varphi|$, is the maximum pure part of $\varphi$: $|\varphi| = \max\{ p \in \Sigma | p \preceq \varphi \land p\text{ is pure} \}$.

	A tuple $(\Sigma, \oplus, u, C, \mathit{Vars}, \sigma)$ (where $\mathit{C}: \Sigma \rightarrow \Sigma$, $\mathit{Vars}$ is a set of variable names,
	and $\sigma: \Sigma \rightarrow \mathcal{P}(\mathit{Vars})$ represents the set of variables declared in a state)
	is a \textbf{separation algebra with a store} $\sigma$ if and only if,
	for all states $a, b, \varphi \in \Sigma$:
	\begin{enumerate}
		\item $(\Sigma, \oplus, u)$ is a partial commutative monoid. \label{axiom:monoid}
		\item $\{ p | p \in \Sigma \land p \text{ is pure} \land p \preceq \varphi \}$ is finite. \label{axiom:finite}
		\item $\varphi = |\varphi| \oplus a \land |a| = u \Longleftrightarrow a = C(\varphi)$ (unique decomposition of a state) \label{axiom:decompo}
		\item $C(\varphi) \oplus a = C(\varphi) \oplus b \Longrightarrow a = b$ ($\oplus$ is partially cancellative) \label{axiom:cancellative}
		\item $a \oplus b = u \Longrightarrow a = u$ (no resource created out of an empty resource) \label{axiom:positivity}
		\item Store properties: \label{axiom:store}
			\begin{enumerate}
				\item $\sigma(C(\varphi)) = \varnothing \land \sigma(a \oplus b) = \sigma(a) \cup \sigma(b)$ (stores are pure and can be added) \label{axiom:store_add}
				\item $\forall x \in \sigma(\varphi). (\exists c \in \Sigma \ldotp \sigma(c) = \{x\} \land c \preceq \varphi )$ (stores with one variable) \label{axiom:store_single}
				\item $a \# b \land a \text{ is pure} \land
					\sigma(a) \subseteq \sigma(b) \Longrightarrow a \preceq b$ (compatible stores) \label{axiom:store_incl}
				\item $\sigma(a) \cap \sigma(b) = \varnothing \land a \text{ is pure} \Longrightarrow a \# b$ (disjoint stores are compatible) \label{axiom:store_disjoint}
			\end{enumerate}
	\end{enumerate}
\end{definition}

In this algebra, any state $\varphi$ can be uniquely decomposed into the sum of its impure resources $C(\varphi)$ and its core $|\varphi|$ (Axiom~\ref{axiom:decompo}).
Axiom~\ref{axiom:monoid} is similar to the definition of a separation algebra in~\cite{Calcagno2007}, without the cancellative axiom (which corresponds to our Axiom~\ref{axiom:cancellative}).
The core of $\varphi$ ($|\varphi|$) is well-defined thanks to Axiom~\ref{axiom:finite}, and contains the store of local variables from $\varphi$, whose domain of definition is given by $\sigma(\varphi)$.
$\sigma(\varphi)$ represents the set of variables that are declared.
The operation $\oplus$ is cancellative for states with only impure resources (Axiom~\ref{axiom:cancellative}).
No resources can be created out of an empty resource (Axiom~\ref{axiom:positivity}).
$C(\varphi)$ has an empty store, and adding two states results in the union of their stores (Axiom~\ref{axiom:store_add}).
Axiom~\ref{axiom:store_single} specifies that it is always possible to reduce a local store to one variable.
Finally, Axioms~\ref{axiom:store_incl} and~\ref{axiom:store_disjoint} ensure that pure states only contain pure resources, and specify the compatibilities of stores.
Two states can be added only if they agree on the values of local variables they both define (Axiom~\ref{axiom:store_incl}).
If the addition is defined, the resulting state is the state that defines all local variables defined by one state or the other (Axiom~\ref{axiom:store_add}), and whose resources are the combination of the resources held by both states (implied by the axioms).
This separation algebra captures a wide range of separation logics and features, including fractional permissions, predicates, and magic wands.

\subsection{Abstract Verification Language}

We define a parametric verification language, whose states are elements of a fixed separation algebra $(\Sigma, \oplus, u, C, \mathit{Vars}, \sigma)$.
The statements of this language are:
	\begin{align*}
		S \definedas \; &S \seq S \; | \; \NDIf{S}{S} 
		\; | \; \whilenoinv{A}{S}
		\; | \; \vecto{V} := m(\vecto{V})
                \; | \; \cskip\\
		\; &| \; \bcmd{assume} A
		\; | \; \bcmd{assert} A
		\; | \; \bcmd{inhale} A 
		\; | \; \bcmd{exhale} A
		\; | \; \bcmd{var} \vecto{V} \; | \; \bcmd{havoc} \vecto{V}
		\; | \; \bcmd{custom} O 
	\end{align*}
where $A$ represents assertions, $\vecto{V}$ lists of elements of $\mathit{Vars}$ (variable names), and $O$ is a type representing any other statement.
Most statements of this language are standard and have the usual semantics.
The $\mathbf{inhale}$ and $\mathbf{exhale}$ statements explicitly manipulate resources. 
Inhaling an assertion adds the resources specified in the assertion to the program state, and renders the execution infeasible if the state is not compatible with the assertion.
Exhaling an assertion removes the resources specified in the assertion from the state, and fails if no state smaller than the current program state satisfies
the assertion. 
$\mathbf{inhale}$ and $\mathbf{exhale}$ statements are used in particular to define the semantics of method calls below.
$\mathbf{havoc}$ non-deterministically assigns new values to a list of variables.
$\bcmd{assume} A$ stops an execution if the current program state does not satisfy $A$ (i.e., the rest of the program is trivially verified),
and behaves like $\mathbf{skip}$ if the current program state satisfies $A$.
Our $\mathbf{assert}$ statement is intrinsically \emph{intuitionistic} (\wrt{} the induced order), since we only use $\mathbf{assert}$ when inlining with partial annotations,
to assert annotations of the form $P * \truesym{}$):
$\bcmd{assert} A$ raises an error if no state smaller than the current program state satisfies $A$, 
and behaves like $\mathbf{skip}$ otherwise.\footnote{It would be no problem to also add an $\mathbf{assert}$ that behaves as in classical SL.}
Note that both statements fail if the assertion $A$ is not compatible with the current program state.
Finally, $\bcmd{custom} O$ represents any other statements, for example a variable assignment.

\paragraph{Assertions.}
Assertions in separation algebras are usually represented as functions from states to booleans~\cite{Reynolds2002,Yang2002,Calcagno2007,OHearn2009}.
To capture automatic SL verifiers,
we add the possibility that an assertion is not well-defined in a state.
For example, the assertion \vipercode{x.f == 5} is well-defined only in states that own the resource \vipercode{x.f}.
Assertions can thus also result in $\mathit{Error}$.

\begin{definition}
	An \textbf{assertion} is a function $\Sigma \rightarrow \{ \top , \bot, \mathit{Error} \}$.
\end{definition}


To define the semantics of the language, we use the function
$sem_\mathcal{A}: \Sigma \times \mathit{Stmt} \rightarrow \mathcal{P}(\Sigma) \cup \{ \bot \}$,
where $\mathcal{A}$ is an annotation containing pre- and postconditions for methods and invariants for loops.
For a state $\varphi \in \Sigma$ and a statement $s$, $sem_\mathcal{A}(\varphi, s) = \bot$ if executing $s$ in the state $\varphi$ fails,
otherwise $sem_\mathcal{A}(\varphi, s)$ is the set of states resulting from the execution of $s$ in the state $\varphi$.
We also define $ver_\mathcal{A}(\varphi, s) \definedas{} \left( sem_\mathcal{A}(\varphi, s) \neq \bot \right)$.

%

Given an assertion $P$, we write $\langle P \rangle$ to refer to the set of states in which $P$ holds, i.e., $\langle P \rangle \definedas \{ \sigma \mid P(\sigma) = \top \}$.
Inhaling $P$ adds to the program state all states contained in $\langle P \rangle$ that are compatible with the program state,
while exhaling $P$ removes \emph{some of} those states (based on the verifier heuristics).
To express the semantics of $\mathbf{inhale}$, we first need to lift the addition and the order relation to sets of states:
\begin{definition}\textbf{Addition and partial order of sets of states.}\label{def:order_sets}
	For $T, U \subseteq \Sigma$,\\
	$T \oplus U \definedas \{ \varphi_1 \oplus \varphi_2 | \varphi_1 \in T \land \varphi_2 \in U \land \varphi_1 \# \varphi_2 \}$ and
	$U \preceq T \Longleftrightarrow (\forall \varphi_1 \in T. \exists \varphi_2 \in U. \varphi_2 \preceq \varphi_1)$.
\end{definition}

When inhaling resources, an automatic verifier does not have any choice.
Consider as an example the existentially-quantified assertion \code{Q(?x)} from \figref{fig:viper-intro-wildcard}.
Inhaling this assertion means obtaining ownership of \code{Q(x)} for some arbitrary \code{x}, and so the verifier must consider all possibilities for \code{x}.
On the other hand, an automatic verifier \emph{does} have the choice when exhaling:
If the current state owns both \code{Q(a)} and \code{Q(b)}, to exhale \code{Q(?x)}, the verifier has the choice to remove either \code{Q(a)} or \code{Q(b)}.

Therefore, to model the heuristics used by the verifier when exhaling, the semantics is parameterized by a function $\mathit{heur}$ that 
maps a state $\varphi$ and a set of states satisfying an assertion $P$ ($\langle P \rangle$) to a non-empty subset of $\langle P \rangle$.
$\mathit{heur}(\varphi, \langle P \rangle)$ represents the choice made by the verifier when exhaling $P$ in the state $\varphi$:
It will only remove from $\varphi$ the states contained in $\mathit{heur}(\varphi, \langle P \rangle)$.
This captures the behavior of \grasshopper{}, \verifast{}, and \Viper{}.
For the semantics of exhale to soundly overapproximate SL's frame rule, we only require that if there is a state smaller than $\varphi$ in $\langle P \rangle$,
then there should also be a state smaller than $\varphi$ in $\mathit{heur}(\varphi, \langle P \rangle)$.
The semantics of inhale and exhale statements is defined as follows:

\begin{definition}\textbf{Semantics of inhale and exhale statements.}\label{def:inhale}\\
%
	$
		sem_{\mathcal{A}}(\varphi, \bcmd{inhale} P) \definedas
		\begin{cases}
			\bot &\text{if } P(\varphi) = \mathit{Error} \\
			\{ \varphi \} \oplus \langle P \rangle &\text{otherwise}
		\end{cases}
	$\\
	$	sem_{\mathcal{A}}(\varphi, \bcmd{exhale} P) \definedas
		\begin{cases}
			\bot &\text{if } P(\varphi) = \mathit{Error} \lor \not\exists i \in \langle P \rangle \ldotp i \preceq \varphi \\
			\{|i| \oplus r | i \in \mathit{heur}(\varphi, \langle P \rangle) \land \varphi = i \oplus r \}
			&\text{otherwise}
		\end{cases}
		$
\end{definition}

$\bcmd{inhale} P$ succeeds if $P$ is well-defined in the program state, and adds the resources specified by $P$ to the state.
$\bcmd{exhale} P$ verifies if $P$ is well-defined in the current program state and $P$ holds in a state smaller than the current program state, in which case
it removes from the state the impure resources specified by $P$\, following the verifier's heuristics (recall that $|i|$ represents the pure part of the state $i$, \ie{} the values of local variables).
As an example, consider the assertion $P \definedas{} \fpointsto{\code{x.f}}{5}{0.5}$, interpreted in a classical SL way.
In this case, $\langle P \rangle$ is a singleton, which contains the state with half ownership of \vipercode{x.f} and where \vipercode{x.f = 5}.
Inhaling $P$ results in an empty set of states if more than half ownership was already held to \vipercode{x.f}, or if some ownership of \vipercode{x.f} was already held and its value was different from $5$.
In all other cases, inhaling $P$ results in the current state, with half more ownership of \vipercode{x.f} and where its value is $5$.
Moreover, note that $sem_\mathcal{A}(\varphi, \bcmd{exhale} P)$ can result in a set with more than one element, if there exist different decompositions of $\varphi$ into $i \oplus r$,
which is the case when $P$ contains, for example, existentially-quantified permission amounts.

Finally, in order to define the semantics of the statements $\mathbf{var}$ (declares a list of variables) and $\mathbf{havoc}$ (non-deterministically assigns values to a list of variables),
we first define two auxiliary functions, $h$ and $\overline{h}_{\vecto{V}}$.
$h(\vecto{V})$ is the set of all pure states (that is states that contain variables but no impure resources) that exactly define the variables of the set $\vecto{V}$.
$\overline{h}_{\vecto{V}}(\varphi)$ is the same state as $\varphi$, except that all variables of $\vecto{V}$ have been removed.

\begin{definition}\textbf{Two auxiliary functions needed to define the semantics of declaring and havocing variables.}
\[
	h(\vecto{V}) \definedas \{ \varphi \in \Sigma. |\varphi| = \varphi \land \sigma(\varphi) = \vecto{V} \}
\]
\[
	\overline{h}_{\vecto{V}}(\varphi) \definedas \text{the unique } b \in \Sigma \text{ such that } b \preceq \varphi \land \sigma(b) = \sigma(\varphi) \setminus \vecto{V} \land |b| = b
\]
\end{definition}

We say that \emph{a statement $s$ verifies modularly in a state $\varphi$ w.r.t\ an annotation $\mathcal{A}$}, denoted by $\mathit{ver}_\mathcal{A}(\varphi, s)$, if $\mathit{sem}_\mathcal{A}(\varphi, s) \neq \bot$.
The semantics of our abstract language is defined as follows:

\begin{definition}\textbf{Semantics} \\
	\label{def:complete_semantics}

	\noindent \textbf{Method calls.}
	Assume the method $m$ has arguments $\vecto{args}$,  return variables $\vecto{rets}$ (where $\vecto{args}$ and $\vecto{rets}$ are lists of variables), and precondition $P$ and  postcondition $Q$ in $\mathcal{A}$.
	Let $P' \definedas P[\vecto{x}/\vecto{args}]$ and $Q' \definedas Q[\vecto{x}/\vecto{args}][\vecto{y}/\vecto{rets}]$.
	\begin{align*}
		&sem_\mathcal{A}(\varphi, \vecto{y} := m(\vecto{x})) \definedas
		\begin{cases}
			\bot \text{ if } \vecto{x} \cup \vecto{y} \not\subseteq \sigma(\varphi) \\
			sem_\mathcal{A}(\varphi, \bcmd{exhale} P' \seq \bcmd{havoc} \vecto{y} \seq \bcmd{inhale} Q') \text{ otherwise}
		\end{cases}
	\end{align*}

	\noindent \textbf{Loops.}
	Let $w \definedas{} \bwhilenoinv$, $\vecto{l} \definedas{} \mathit{modif}(s)\cap \sigma(\varphi)$, and let $I$ be the invariant for $w$ in annotation $\mathcal{A}$.
	\begin{align*}
		&sem_\mathcal{A}(\varphi, w) \definedas
 		\begin{cases}
 			\bot \text{ if } \lnot ver_\mathcal{A}(\{|\varphi|\}, \bcmd{havoc} \vecto{l} \seq \bcmd{inhale} I \seq \bcmd{assume} b \seq s \seq \bcmd{exhale} I) \\
 			sem_\mathcal{A}(\varphi, \bcmd{exhale} I \seq \bcmd{havoc} \vecto{l} \seq \bcmd{inhale} I \seq \bcmd{assume} \lnot b) \text{ otherwise}
 		\end{cases}
	\end{align*}

	\noindent
	\textbf{Other statements.}
	\begin{align*}
		&sem_\mathcal{A}(\varphi, \NDIf{s_1}{s_2}) &&\definedas
		\begin{cases}
			\bot \text{ if } sem_\mathcal{A}(\varphi, s_1) = \bot \lor sem_\mathcal{A}(\varphi, s_2) = \bot \\
			sem_\mathcal{A}(\varphi, s_1) \cup sem_\mathcal{A}(\varphi, s_2) \text{ otherwise}
		\end{cases} \\
		&sem_\mathcal{A}(\varphi, s_1 \seq s_2) &&\definedas
		\begin{cases}
			\bot \text{ if } A = \bot \lor (\exists a \in A. sem_\mathcal{A}(a, s_2) = \bot) \\
			\bigcup_{a \in A} sem_\mathcal{A}(a, s_2) \text{ otherwise}
		\end{cases} \tag{where $A \definedas sem_\mathcal{A}(\varphi, s_1)$} \\
		&sem_\mathcal{A}(\varphi, \bcmd{assume} A) &&\definedas
		\begin{cases}
			\bot &\text{ if } A(\varphi) = Error \\
			\varnothing &\text{ if } A(\varphi) = \bot \\
			\{ \varphi \} &\text{ if } A(\varphi) = \top \\
		\end{cases} \\
		&sem_\mathcal{A}(\varphi, \bcmd{assert} A) &&\definedas
		\begin{cases}
			\{ \varphi \} &\text{ if } A(\varphi) \neq \mathit{Error} \land \exists i \in \langle P \rangle \ldotp i \preceq \varphi \\
			\bot &\text{ otherwise }
		\end{cases} \\
		&sem_\mathcal{A}(\varphi, \bcmd{var} \vecto{V}) &&\definedas
		\begin{cases}
			\bot &\text{ if } \vecto{V} \cap \sigma(\varphi) \neq \varnothing \\
			\{ \varphi \} \oplus h(\vecto{V})  &\text{ otherwise}
		\end{cases} \\
		&sem_\mathcal{A}(\varphi, \bcmd{havoc} \vecto{V}) &&\definedas
		\begin{cases}
			\bot &\text{ if } \vecto{V} \nsubseteq \sigma(\varphi) \\
			\{ \overline{h}_{\vecto{V}}(\varphi) \} \oplus h(\vecto{V}) &\text{ otherwise}
		\end{cases} \\
		&sem_\mathcal{A}(\varphi, \bcmd{skip}) &&\definedas \{ \varphi \}
	\end{align*}
\end{definition}

The semantics of method calls requires that the corresponding method has (potentially trivial) pre- and postconditions in $\mathcal{A}$ (and analogously for loops).

In the semantics of the loop,
$\vecto{l}$ contains all declared local variables modified by the loop.
The loop body verifies starting from a state that contains no resources and in which all local variables modified by the loop have arbitrary values.
The invariant is then inhaled, the loop guard assumed, the loop body $s$ is executed, and finally the loop invariant is exhaled.
If this verifies, the semantics of the loop amounts to exhaling the loop invariant, assigning arbitrary values to local variables modified by the loop, inhaling the invariant, and assuming the negation of the loop guard.

Declaring a list of variables verifies if and only if these variables are not already defined in the program state.
In this case, the resulting states are the states that contain all resources from before, plus where the specified variables are defined (with any value).
Havocing a list of variables verifies if and only if these variables are already defined in the program state.
In this case, we simply forget the values of these variables.

\section{Preserving Inlining}~\label{sec:semantics_loops}

\begin{definition}\textbf{Inlining definition (ignoring renaming issues).}
	\begin{align*}
		\mathit{inl}^0_M(\vecto{y} := m(\vecto{x})) &\definedas \bcmd{assume} \bot \\
		\mathit{inl}^{n+1}_M(\vecto{y} := m(\vecto{x})) &\definedas \mathit{inl}^n_M(s_m) \\
		\mathit{inl}^0_M(\bwhilenoinv) &\definedas \bcmd{assume} \lnot b \\
		\mathit{inl}^{n+1}_M(\bwhilenoinv) &\definedas \bcmd{if} (b) \; \{\mathit{inl}^n_M(s) \seq \;\mathit{inl}^n_M(\bwhilenoinv)\} \\
		\mathit{inl}^n_M(s_1 \seq s_2) &\definedas \mathit{inl}^n_M(s_1) \seq \mathit{inl}^n_M(s_2) \\
		\mathit{inl}^n_M(\NDIf{s_1}{s_2}) &\definedas \NDIf{\mathit{inl}^n_M(s_1)}{\mathit{inl}^n_M(s_2)} \\
		\mathit{inl}^n_M(s) &\definedas s \tag{in all other cases}
	\end{align*}
where $\bcmd{if} (b) \; \{s\}$ is syntactic sugar for $\NDIf{\bcmd{assume} b \seq s}{\bcmd{assume} \lnot b}$
and $s_m$ is the body of method $m \in M$.
\end{definition}

When the remaining inlining bound $n$ has reached $0$, we make no further loop iterations.
Otherwise, inlining a loop amounts to inlining its first iteration and then inlining the rest of the loop.
Note that we use one global inlining bound here. It would also be possible to use separate bounds for the call depth and for each loop, respectively.

\begin{definition}\textbf{\Soundness{} condition (ignoring renaming issues).}
	\begin{align*}
		PC^n_M(T, s) &\Longleftrightarrow \mathit{mono}_{\emptyannot}(T, s) \tag{if $s$ does not contain any method call or loop} \\
		PC_M^n(T, \NDIf{s_1}{s_2}) &\Longleftrightarrow PC^n_M(T, s_1) \land PC^n_M(T, s_2) \\
		PC_M^n(T, s_1 \seq s_2) &\Longleftrightarrow PC^n_M(T, s_1) \land PC^n_M(\overline{\mathit{sem}}_{\emptyannot}(T, \mathit{inl}_M^n(s_1)), s_2) \\
		PC_M^0(T, \vecto{y} := m(\vecto{x})) &\Longleftrightarrow \top \\
		PC_M^{n+1}(T, \vecto{y} := m(\vecto{x})) &\Longleftrightarrow
			\mathit{framing}_M(T, \mathit{inl}_M^n(s_m))
			\land PC^n_M(T, s_m) \\
 		PC_M^0(T, \bwhilenoinv) &\Longleftrightarrow \mathit{mono}_{\emptyannot}(T, \bcmd{assume} \lnot b) \\
 		PC_M^{n+1}(T, \bwhilenoinv) &\Longleftrightarrow \\
 			&(\forall U \subseteq \Sigma. \mathit{mono}_{\emptyannot}(U, \bcmd{assume} b) \land \mathit{mono}_{\emptyannot}(U, \bcmd{assume} \lnot b))  \\
 			&\land
 			\mathit{framing}_{\emptyannot}(T_b, \mathit{inl}_M^n(s)) \land
 			PC_M^n(T_b, s) \land PC_M^n(T'', \bwhilenoinv)
 	\end{align*}
 	where $\emptyannot{}$ is the empty annotation, $s_m$ is the body of method $m \in M$,\\
	$\overline{\mathit{sem}}_{\mathcal{A}}(T, s) \mapsto \bigcup_{\varphi \in \Sigma | (\exists \varphi' \in T. \varphi \preceq \varphi') \land \mathit{ver}_{\mathcal{A}}(\varphi, s)}( \mathit{sem}_{\mathcal{A}}(\varphi, s) )$,
	\\
	$T_b \definedas \{ \varphi | \varphi \in T \land b(\varphi) = \top \}$, and
	$T'' \definedas \overline{\mathit{sem}}_{\emptyannot{}}(T_b, \mathit{inl}_M^n(s))$.
\label{def:appendix-complete-sc}
\end{definition}

In the sequential composition and loop cases, we need to introduce the auxiliary function $\overline{sem}$,
which applies the $\mathit{sem}$ function to the states of the transitive closure of $T$ in which $s$ verifies.
This transitive closure is required to compute the right resource bound for the framing and monotonicity properties,
in order to ensure \sound{} inlining.
As shown  in Definition~\ref{def:appendix-complete-sc}, after the final iteration of the loop has been inlined, the \soundness{} condition requires $\mathit{mono}_{\epsilon}(T, \bcmd{assume} \lnot b)$, which reflects that inlining cuts off all executions that still satisfy the loop condition after the bound is reached ($n=0$).
Finally, in the loop case where the bound has not been reached ($n>0$), 
the \soundness{} condition demands that inlined loop iterations are framing. Moreover, $\bcmd{assume} b$ and $\bcmd{assume} \lnot b$ must be unboundedly mono in this case (where $b$ is the loop condition).
This simply means that if the assertion loop condition is well-defined ($\neq \mathit{Error}$) for a state $\varphi$ then the loop condition will also be defined for any state larger than $\varphi$, and will have the same value.
For example, the assertions \vipercode{i < n} and \vipercode{x.f != 0} both satisfy this condition, whereas the assertion \vipercode{perm(x.f) >= 1/2} does not.
Therefore, if \vipercode{perm(x.f) >= 1/2} is used in a loop condition, then the \soundness{} condition does not hold. Intuitively, this is the case, because owning different fractional ownership of \code{x.f} leads to \vipercode{perm(x.f) >= 1/2} being evaluated differently, which can lead to \unsound{} inlining (since different branches may be taken in the inlined and original programs).

\subsection{Partial annotations and loops}
For inlining to take partial annotations into account with programs that contain loops in addition to calls, the syntactic transformation $\mathit{assertAnnot}$ defined in~\secref{sub:inl-partial-annot} that is applied before inlining the program additionally must take invariants into account.
If a program $(s,M)$ contains loops, then $\mathit{assertAnnot}_\mathcal{A}(s,M)$ is defined as in~\secref{sub:inl-partial-annot} except that $\mathit{assertAnnotStmt}$ is lifted to handle loops as we show next.

\begin{definition}
	\textbf{The $\mathit{assertAnnotStmt}$ syntactic statement transformation.}\\

	\noindent \textbf{MethodCalls.}
	Assume the method $m$ has arguments $\vecto{args}$,  return variables $\vecto{rets}$ (where $\vecto{args}$ and $\vecto{rets}$ are lists of variables), and precondition $P$ and  postcondition $Q$ in $\mathcal{A}$.
	Let $P' \definedas P[\vecto{x}/\vecto{args}]$ and $Q' \definedas Q[\vecto{x}/\vecto{args}][\vecto{y}/\vecto{rets}]$.

	\begin{align*}
		\mathit{assertAnnotStmt}_\mathcal{A}(\vecto{y} := m(\vecto{x})) \definedas \bcmd{assert} P' * \truesym{} \seq \vecto{y} := m(\vecto{x}) \seq \bcmd{assert} Q' * \truesym{}
	\end{align*}

	\noindent \textbf{Loops.}
	Assume the while loop below has invariant $I$ in $\mathcal{A}$
	\begin{align*}
		&\mathit{assertAnnotStmt}_\mathcal{A}(\bwhilenoinv) \definedas \\
		&\quad \bcmd{assert} I * \truesym{} \seq \whilenoinv{b}{\bcmd{assert} I * \truesym{} \seq
		\mathit{assertAnnotStmt}_\mathcal{A}(s)
		\seq \bcmd{assert} I * \truesym{}} \seq \bcmd{assert} I * \truesym{}
	\end{align*}

	\noindent \textbf{Other statements.}
	\begin{align*}
		&\mathit{assertAnnotStmt}_\mathcal{A}(s_1 \seq s_2) \definedas \mathit{assertAnnotStmt}_\mathcal{A}(s_1) \seq \mathit{assertAnnotStmt}_\mathcal{A}(s_2) \\
		&\mathit{assertAnnotStmt}_\mathcal{A}(\NDIf{s_1}{s_2}) \definedas \\
		& \quad \NDIf{\mathit{assertAnnot}_\mathcal{A}(s_1)}{\mathit{assertAnnot}_\mathcal{A}(s_2)} \\
		&\mathit{assertAnnotStmt}_\mathcal{A}(s) \definedas s \tag{in all other cases}
	\end{align*}
	\label{def:assertAnnotWithLoops}
\end{definition}

\section{Mechanization in \isabelle{}}~\label{app:mechanization}

Our mechanization closely follows the formal definitions given in \secref{sec:soundness}, \appref{app:soundness}, and \appref{sec:semantics_loops}, with the following notables differences:
\begin{itemize}
    \item Annotations in the mechanization are part of the program, whereas they are separate from the program in this paper. This difference simplifies the mechanization, but the results are the same as presented here.
    \item We chose to ignore renaming issues in our presentation, but the mechanization handles them. Moreover, the mechanization of framing includes the side condition on local variables.
\end{itemize}

%
%

\section{Dual Result to Preserving Inlining}
\label{app:completeness}

We have proved in \secref{sec:soundness} that inlining is \sound{} under the \soundness{} condition (Theorem~\ref{thm:soundness}): Any error found in the inlined program corresponds to a true error in the original program. In other words, the \soundness{} condition implies that inlining does not introduce false positives (spurious errors). In this section, we prove a dual result: If the inlined program verifies, then there is no true error for bounded executions of the original program.
In other words, inlining does not introduce false negatives in bounded executions.
We call this result \emph{completeness of inlining} (or simply \emph{completeness}).

To prove completeness, we first define a bounded semantics of the original program by instrumenting it with ghost counters and  assumptions that reflect the inlining bound. Moreover, we define the concept of a \emph{maximally-annotated} program, whose method specifications and loop invariants are precise, and show how any program can be instrumented so that it can be maximally-annotated.
We then prove on paper that if the inlined version of a program verifies, then the original maximally-annotated program verifies with the bounded semantics (see~\cite{artifact} for the full details of the proof; we provide the key ideas in this section).

\subsection{Bounded semantics}
Since inlining is only up to a bound, it is clear that the verification of the inlined program can only inform us on similarly bounded executions of the original program. The example in \figref{fig:bounded-semantics} illustrates this limitation of bounded verification. The original program (on the left) fails to verify since \code{n} is unconstrained. However, the program inlined with a bound of $2$ (shown in the middle) verifies, since all executions that perform more than two iterations are infeasible.

\begin{figure}
\begin{minipage}[t]{0.28\textwidth}
\begin{viper2}
method m(n: Int)
{
  var i: Int := 0
  while (i < n) {
    i := i + 1
  }
  assert n <= 2
}\end{viper2}
\end{minipage}
\begin{minipage}[t]{0.31\textwidth}\begin{viper2}
method inl(n: Int)
{
  var i: Int := 0
  if (i < n) {
    i := i + 1
    if (i < n) {
      i := i + 1
      assume !(i < n)
    }
  }
  assert n <= 2
}\end{viper2}\end{minipage}
\begin{minipage}[t]{0.4\textwidth}
\begin{viper2}
method bounded(n: Int)
{
  var i: Int := 0
  var dep: Int := $b$
  while (i < n && dep > 0) {
    dep := dep - 1
    i := i + 1
  }
  assume !(i < n)
  assert n <= 2
}\end{viper2}\end{minipage}
\caption{The original program on the left, the same program inlined with a bound of $2$ in the middle, and the bounded semantics instrumentation on the right.}
\label{fig:bounded-semantics}
\end{figure}

We deal with this fundamental limitation of bounded verification by defining a \emph{bounded semantics}, which is similar to the normal semantics, but executions that go beyond the inlining bound are infeasible. We define the bounded semantics as a program instrumentation with ghost counters and assume statements that stop executions exceeding the bound. For each loop, we introduce a ghost variable \code{dep}, which is initialized to the inlining bound $b$ and decreases in each iteration. The loop stops when \code{dep} reaches 0. The instrumentation of methods is analogous, with the ghost counter  passed as an argument.
The program on the right of \figref{fig:bounded-semantics} shows this instrumentation for the original program on the left. For the bound $b=2$, the instrumented program verifies with the loop invariant \code{$0 \le$ dep $\le 2 \land \phantom{}$ dep + i = 2}.

\subsection{Maximally-annotated programs}

A maximal annotation for a program w.r.t.\ an inlining bound $n$ is an annotation that precisely captures the program states \emph{of the bounded semantics} before and after loop bodies as well as before and after method calls. In particular, these annotations capture all resources in those states such that reasoning about loops and calls does not require framing. Consequently, verification of a  maximally-annotated program with the bounded semantics resembles verification of the inlined program, where the maximal amount of resources is available at each point.

To obtain a maximal annotation for a program, we instrument its bounded semantics with further ghost code such that all method calls and loop iterations that will be inlined are unambiguously indexed.
More precisely, each method call that is inlined can be uniquely identified using the arguments of the method call, and 
each loop iteration of the same loop can be identified using some variables.
Since there is a finite number of loop iterations and method calls that are inlined (for a given bound), any program can be instrumented to satisfy this condition.

We then assume that the assertion language is expressive enough (possibly after instrumenting the program with more ghost code) to write loop invariants and method pre- and postconditions that precisely capture the sets of states at all program points in the bounded semantics that correspond to boundaries of loop iterations and method calls. In particular, these annotations must precisely capture all impure resources as well as the values of local variables.

As an example, consider the initial statement $s_1 \seq m(x) \seq s_2 \seq m(x)$, where the body of $m$ is $s_m$ (we ignore renaming issues here) and $s_1,s_2,s_m$ do not contain any calls or loops.
Inlining this program with bound 1 results in $s_1 \seq s_m \seq s_2 \seq s_m$. We first instrument the original program to add indices: 
$s_1 \seq m(x, 0) \seq s_2 \seq m(x, 1)$, adding a parameter $i$ to method $m$.
Let $P_1$ be an assertion that precisely captures the set of states $\mathit{sem}_{\emptyannot{}}(u, s_1)$ (recall that $u$ is the state containing no resources that an execution begins with), that is $\langle P_1 \rangle = \mathit{sem}_{\emptyannot{}}(u, s_1)$ (assuming the only variables used in the program are $x$ and $i$).
Similarly, let
$Q_1$ precisely capture $\mathit{sem}_{\emptyannot{}}(u, s_1 \seq s_m)$,
$P_2$ precisely capture $\mathit{sem}_{\emptyannot{}}(u, s_1 \seq s_m \seq s_2)$,
and $Q_2$ precisely capture $\mathit{sem}_{\emptyannot{}}(u, s_1 \seq s_m \seq s_2 \seq s_m)$.
Using these assertions, we can annotate the method $m$ with the precondition $(i = 0 \lor i = 1) \land (i = 0 \Rightarrow P_1) \land (i = 1 \Rightarrow P_2)$,
and with the postcondition $(i = 0 \Rightarrow Q_1) \land (i = 1 \Rightarrow Q_2)$.
This annotation is a maximal annotation for this program.

By applying the same reasoning to loops, we can construct a maximal annotation for any program. Our requirement that this maximal annotation precisely captures impure resources corresponds to a \emph{supported} predicate in the sense of~\cite{OHearn2009}.
More precisely, any assertion $P$ from a maximal annotation satisfies the following:
$\forall \varphi \in \langle P \rangle. \mathit{sem}_\mathcal{A}(\varphi, \bcmd{exhale} P) = \{ |\varphi| \}$.

\subsection{Completeness theorem and proof}

We can now express and prove our completeness result:
If a program is maximally-annotated and its inlined version verifies then this annotated program verifies with the bounded semantics.

\begin{theorem}\textbf{Completeness of inlining.}
	\label{thm:completeness}
	If
	\begin{enumerate}
		\item the program $(s, M)$ is well-formed,
		\item the program $(s, M)$ has a maximal annotation $\mathcal{A}$ w.r.t.\ inlining bound $n$, and
		\item the inlined program verifies: $ver_{\emptyannot}(u, inl_M^n(s))$
	\end{enumerate}

	\noindent
	then program $(s, M)$ verifies modularly with the bounded semantics w.r.t. $\mathcal{A}$:
	\begin{enumerate}
		\item The initial statement $s$ verifies modularly w.r.t. the maximal annotation: $\mathit{verB}_\mathcal{A}(\delta_n, s)$ and
		\item all methods of $M$ verify modularly w.r.t. $\mathcal{A}$
	\end{enumerate}
\end{theorem}

\noindent
The function $\mathit{verB}$ (similar to the function $\mathit{ver}$ from \secref{sec:soundness}) represents the bounded verification,
and $\delta_n$ is the state with no impure resources and only one variable: the variable \code{dep} with value $n$, which stores the inlining bound.
The complete proof of this theorem is available online~\cite{artifact}.

\section{Syntactic Mono and Framing}~\label{app:syntactic-cond}

The syntactic mono and framing properties are based on the simple observation that most features in programs processed by automatic separation logic verifiers are both mono and framing. If a program uses only those features, inlining is \sound{}.
We focus here on features of programs processed by \grasshopper{}, \verifast{}, and \Viper{}, but an adaptation to other verifiers should be straightforward.

\paragraph{Imprecise assertions.}
As explained in \secref{sec:problem}, proof search strategies for proving imprecise assertions are often automated with heuristics that depend on the resources owned.
As a result, \unsound{} inlining can occur in programs where imprecise assertions must be proved and the resulting resource used as a witness is given away.
Our syntactic condition rejects programs that use features where this scenario could occur.
This means rejecting programs where imprecise assertions may occur, for example, in the following cases:
\begin{itemize}
\item The imprecise assertion occurs in the precondition of a bodyless library method (possible in \grasshopper{}, \verifast{}, and \Viper{}).
\item Existential parameters are bound in the precondition of a method $m$. This case can lead to \unsound{} inlining when inlining with partial annotations in conjunction with ghost code that relies on how those parameters are instantiated (possible in \verifast{}).
\item The imprecise assertion occurs in the body of a predicate and there is a ghost operation that closes this predicate (possible in \verifast{} and \Viper{}).
\item The imprecise assertion is explicitly given up via an exhale operation (possible in \Viper{}).
\item The imprecise assertion is part of ghost code (possible in \verifast{}).
\end{itemize}

The syntactic condition classifies an assertion as imprecise based on the syntax of the assertion, which, in general, leads to an overapproximation. 
For example, an assertion is characterized as imprecise, if it is a disjunction (\grasshopper{}), contains an existential quantification of fractional ownership (\verifast{}, \Viper{}) or contains existentially quantified predicate parameters (\grasshopper{}, \verifast{}).

\paragraph{Resource introspection and exact bounds.}
\Viper{} supports the \code{perm} resource introspection feature to inspect the currently owned features.
The feature is used by multiple verifiers to encode proof search algorithms and can lead to \unsound{} inlining (as shown in~\secref{sec:problem}).
The syntactic condition rejects all \Viper{} programs where \code{perm} appears.
For verifiers that work via translation to \Viper{}, their source programs are rejected if their Viper translations use \code{perm}.

Both \grasshopper{} and \Viper{} support the statement \code{assume A}, where \code{A} can be an arbitrary assertion.
Since such a statement renders executions that have less resources than \code{A} infeasible, but not those with executions with at least as many resources as specified by \code{A}, they are potentially not output monotonic, thus our syntactic condition also rejects them.
Finally, since \grasshopper{}'s assert statements express exact bounds on resource ownership (reflecting \grasshopper{}'s underlying classical separation logic), they are not safety monotonic, thus our syntactic condition rejects them.

\section{Automatically Checking Structural Framing}~\label{app:structural-framing}

\def\HiLi{\leavevmode\rlap{\hbox to \hsize{\color{yellow!50}\leaders\hrule height .8\baselineskip depth .5ex\hfill}}}

Similarly to structural mono in \secref{sec:automation}, we define structural framing by replacing the existential quantification in \defref{def:framing}
by a universal quantification over a non-empty range.

\begin{definition}\textbf{Structural framing}\footnote{In our \isabelle{} mechanization, we use two different $\mathit{det}$ functions for structural mono and structural framing, but we ignore this aspect here for simplicity. Similarly to structural mono, our mechanization considers $\mathit{det}$ to be a parameter of $\mathit{structFraming}$, whereas in this presentation we instantiate the determinization function $\mathit{det}$ with the one that corresponds to our implementation.} \label{def:struct-framing}
	\begin{align*}
		&\mathit{structFraming}_\mathcal{A}(T, s) \definedas \\
		&(\forall \varphi, r \in \Sigma \ldotp
		\varphi \# r \land
		\varphi \oplus r \ll T
		\land \mathit{ver}_\mathcal{A}(\varphi, s) \Longrightarrow
		\mathit{ver}_\mathcal{A}(\varphi \oplus r, s) \land \\
		&\quad \left(
			\begin{array}{l}
			\forall \varphi_r' \in \mathit{sem}_\mathcal{A}(\varphi \oplus r, s) \ldotp
			(\forall \varphi' \in \determFraming \oplus \{ r \} \ldotp \varphi' \preceq \varphi_r')
			\land \phantom{} \\
			\varnothing \subset \determFraming \oplus \{ r \} \subseteq \mathit{sem}_\mathcal{A}(\varphi, s) \oplus \{ r \}
			\end{array}
		\right)
	\end{align*}
\end{definition}

It is easy to show that structural framing implies framing, as we show next and as we have also proved in \isabelle{}.

\begin{lemma}\label{lemma:struct_framing_implies_framing}\textbf{Structural framing implies framing:}
	$\mathit{structFraming}_\mathcal{A}(T, s) \Longrightarrow \mathit{framing}_\mathcal{A}(T, s)$.
\end{lemma}

\begin{proof}
	Given the similar structure between the definitions of framing (\defref{def:framing}) and structural framing (\defref{def:struct-framing}),
	we simply need to show that (a) $
			\forall \varphi_r' \in \mathit{sem}_\mathcal{A}(\varphi \oplus r, s) \ldotp
			(\forall \varphi' \in \determFraming \oplus \{ r \} \ldotp \varphi' \preceq \varphi_r')
			\land
			\varnothing \subset \determFraming \oplus \{ r \} \subseteq \mathit{sem}_\mathcal{A}(\varphi, s) \oplus \{ r \}
			$
	implies (b) $
	\mathit{sem}_\mathcal{A}(\varphi, s) \oplus \{ r \}
	\preceq
	\mathit{sem}_\mathcal{A}(\varphi \oplus r, s)
	$.
	(b) is equivalent, by definition (\defref{def:order_sets}), to $\forall \varphi'_r \in sem_\mathcal{A}(\varphi \oplus r, s) \ldotp \exists \varphi' \in sem_\mathcal{A}(\varphi, s) \oplus \{r \} \ldotp \varphi' \preceq \varphi'_r$.
	We assume (a), and want to show (b).

	Let $\varphi'_r \in sem_\mathcal{A}(\varphi \oplus r, s)$.
	From (a), we know that $\determFraming \oplus \{ r \}$ is not empty.
	Thus, let $\varphi'$ be any state from $\determFraming \oplus \{ r \}$.
	Then, from (a), $\varphi'$ also belongs to $sem_\mathcal{A}(\varphi, s) \oplus \{ r \}$, and $\varphi' \preceq \varphi_r'$ holds,
	which proves (b), and thus concludes the proof.
\end{proof}

\begin{figure}
	\centering
	\begin{algorithmic}[1]
		\IF{(*)} \label{fline:if}
            \STATE \HiLi Let $\varphi_1$, $\varphi_2, r$ be Viper states s.t.\
			$\varphi_2 = \varphi_1 \oplus r$
			and
			$\varphi_2 \preceq \mathit{currentViperState}$
			\label{fline:states}
			\STATE $(\mathit{exist}, \mathit{currentViperState})  \leftarrow (\top, \varphi_1)$ \label{fline:start1}
			\STATE $\mathit{guardExecs}(s, \mathit{exist})$  \label{fline:middle1}
			\STATE $\varphi'_1 \leftarrow \mathit{currentViperState}$ \label{fline:end1}
			\STATE $\mathit{currentViperState} \leftarrow \varphi_2$ \label{fline:start2}
			\STATE $s$ \label{fline:middle2}
            \STATE \HiLi assert $\mathit{exist} \land \varphi'_1 \oplus r \text{ is well-defined} \land \varphi'_1 \oplus r \preceq \mathit{currentViperState}$ \label{fline:assert}
			\STATE assume $\bot$ \label{fline:stop}
		\ENDIF
\end{algorithmic}
    \caption{Proof obligation, expressed via self-composition, to check if a deterministic statement $s$ is structurally mono \emph{and} framing.
    The modifications from \figref{fig:algo} are highlighted.}
	\label{fig:algo_framing}
\end{figure}

In~\secref{sec:automation}, we show how to check whether a statement is structurally mono (see~\figref{fig:algo}).
\figref{fig:algo_framing} shows the proof obligation generated by our tool to check if a statement is structurally mono \emph{and} framing.
We do not need to check framing independently, since all statements that the \soundness{} condition (\defref{def:sc}) requires to be framing are also required to be mono.

The first modification compared to \figref{fig:algo} is on line~\ref{fline:states}:
On top of constraining $\varphi_2$ to be greater than $\varphi_1$, we also keep track of the frame $r$.
Note that, by definition, the existence of a state $r$ such that $\varphi_2 = \varphi_1 \oplus r$ is equivalent to $\varphi_1 \preceq \varphi_2$.

The second modification is on line~\ref{fline:assert}.
We still need to check that $\mathit{exist}$ holds, for the same reason as explained in \secref{sec:automation}.
Moreover, we need to check a stronger property than $\varphi'_1 \preceq \mathit{currentViperState}$.
Indeed, to prove that the statement $s$ is framing, we need to prove (see \defref{def:framing} and \defref{def:order_sets}) that,
for any $\mathit{currentViperState} \in sem_\mathcal{A}(\varphi_2, s)$, there exists $\varphi'_1 \in sem_\mathcal{A}(\varphi_1, s) \oplus \{ r \}$ such that
$\varphi'_1 \preceq \mathit{currentViperState}$.
If $\mathit{exist}$ holds, then we know that $\varphi'_1$ is indeed an element of $sem_\mathcal{A}(\varphi_1, s)$.
Furthermore, if $\varphi'_1 \oplus r$ is defined, then (by definition) $\varphi'_1 \oplus r$ is an element of $sem_\mathcal{A}(\varphi_1, s) \oplus \{ r \}$.
Thus, if the assertion on line~\ref{fline:assert} holds, we know there exists
a state $\varphi'_1 \in sem_\mathcal{A}(\varphi_1, s) \oplus \{ r \}$ such that $\varphi'_1 \preceq \mathit{currentViperState}$, which proves that $s$ is framing.

\section{Determinization}~\label{app:determinization}

\newcommand{\mathcolorbox}[2]{\colorbox{#1}{$\displaystyle #2$}}

In this section, we first concretely show, via an example on \figref{fig:determinization}, how we instrument a statement so that we can compare two executions of this
statement where non-determinism is resolved in the same way, which is required for our structural mono and framing checks.
We then explain two aspects of how the approach works in general.

\subsection{Determinization on an Example}

\begin{figure}
\begin{minipage}[t]{0.33\textwidth}
	\centering
\begin{viper2}





var b: Bool

if (b) {
  inhale $\fpointsto{\code{x.f}}{\_}{*}$

}
\end{viper2}\end{minipage}\begin{minipage}[t]{0.33\textwidth}
\centering
\begin{viper2}
@\HiLi@var detB: Bool
@\HiLi@var detXF: Int
@\HiLi@var detP: Perm
@\HiLi@assume detP > 0
var b: Bool
@\HiLi{}@detB := b
if (b) {
  inhale $\fpointsto{\code{x.f}}{\_}{\text{\mathcolorbox{yellow}{\mathit{detP}}}}$
@\HiLi{}@  detXF := x.f
}
\end{viper2}\end{minipage}\begin{minipage}[t]{0.33\textwidth}
\centering
\begin{viper2}





var b: Bool
@\HiLi{}@assume b = detB
if (b) {
  inhale $\fpointsto{\code{x.f}}{\_}{\text{\mathcolorbox{yellow}{\mathit{detP}}}}$
@\HiLi{}@  assume x.f = detXF
}
\end{viper2}\end{minipage}
\vspace{-4mm}
    \caption{A simplified example of how determization works with the original statement $s$ (on the left).
    $\fpointsto{\code{x.f}}{\_}{*}$ is equivalent to $\exists k > 0 \ldotp \fpointsto{\code{x.f}}{\_}{k}$,
    thus \code{inhale $\fpointsto{\code{x.f}}{\_}{*}$} non-deterministically adds some fractional non-zero ownership of \code{x.f}
    to the state.
    An instrumentation of the statement $s$ to record non-deterministic choices is shown on the middle (we call it the first execution).
    The instrumentation on the right shows how we ignore executions of $s$ that resolve non-determinism differently (the second execution).
    We assume (in this particular example) that the initial state does not hold any ownership of $x.f$.
    The modifications from the original statement are highlighted in yellow.
    }
\label{fig:determinization}
\end{figure}

Determinization first instruments the original statement to record the non-deterministic choices that are taken in a first execution
(shown in the middle of \figref{fig:determinization}),
then instruments a second time the original statement to stop the execution if non-determinism is resolved in a different way (right of \figref{fig:determinization}).

For simplicity, we ignore here the \textit{guardExecs} transformation (described in \secref{sec:automation}).
If executed in an initial state that does not hold any ownership of $x.f$ and $y.f$,
the execution of the statement $s$ (on the left) has (at most) three non-deterministic choices to make:
\begin{enumerate}
    \item The initial value of the local variable $b$. We record this choice with the variable \code{detB}.
    \item The fractional ownership of $x.f$ that the statement \vipercode{inhale} $\fpointsto{\code{x.f}}{\_}{*}$ adds to the state
        ($\fpointsto{\code{x.f}}{\_}{*}$ is equivalent to $\exists k > 0 \ldotp \fpointsto{\code{x.f}}{\_}{k}$),
        recorded with the variable \code{detP}.
        Moreover, the value of \code{detP} is initially constrained to be any strictly positive rational.
        Thus, for each possible fractional ownership witness there is a corresponding execution that reflects this value in \code{detP}.
    \item Since the execution does not hold any ownership of \code{x.f} before the statement $s$,
    adding ownership also has the effect of non-deterministically assigning an initial value to the heap location \code{x.f}.
    We thus record this initial value with the variable \code{detXF}.
\end{enumerate}
Using these three variables (\code{detB}, \code{detP}, and \code{detXF}),
we ignore executions of $s$ that resolve non-determinism differently, as shown on the right of \figref{fig:determinization}.

It is important to note that all variables used to record non-determinism are assigned at most once.
Thus, if the assignment of such a variable in the first execution is not reached, 
constraining the non-deterministic choice in the second execution $E_2$ does not have any consequence,
since this variable can have all possible values.

\subsection{Determinization in General}

\paragraph{Initial values of heap locations.}
In the previous example, we record (and then constrain) the value of \code{x.f} to be the same in both executions.
This approach is correct only when both executions have no ownership of \code{x.f} before this statement.
In the other cases, recording and constraining the value of \code{x.f} to be the same in both executions could potentially make our automated check ignore some executions,
and thus it would not correspond to the mono or framing property.
That is why we only match initial values of heap location when both executions have no ownership of \code{x.f} before this statement.

\paragraph{Existential fractional ownership amounts.}
One can always match existential fractional ownership amounts that are added to the state (\vipercode{inhale}), as shown in the example.
However, it is not possible in general to perfectly match them when they are removed from the state (\vipercode{exhale}).
As an example, consider:
\begin{itemize}
    \item the statement \vipercode{exhale} $\fpointsto{\code{x.f}}{\_}{*}$ (remove some non-zero fractional ownership of \code{x.f}),
    \item the execution $E_1$ starting from the Viper state $\varphi_1$ with half ownership of \code{x.f},
    \item the execution $E_2$ starting from the Viper state $\varphi_2$ with full ownership of \code{x.f}.
\end{itemize}
There exists an execution in the initial state $\varphi_2$ where the exhale removes $\frac{3}{4}$ ownership.
However, we cannot match it with an execution starting in $\varphi_1$ where the same fraction ($\frac{3}{4}$) of ownership is removed,
since $\varphi_1$ has only half ownership.
In this case, we simply constrain the execution $E_2$ to not end up with less ownership than $E_1$ after the execution of this statement,
if it had not less before the statement.

\section{Evaluation of Examples Taken from Viper's Test Suite}
\label{app:syntactic_evaluation}
\begin{table}[H]
    \centering
    \begin{tabular}{l|c|c|c|c|c|c|c|c|c|c|r}
    Name  & LOC & Ann. & \#Err. & S.Err. & Inl.P. & SC & Str.C. & T [sec]\\
    \hhline{=|=|=|=|=|=|=|=|=|=|=}
    \code{array\_list} & 63 & 27 & 5 & 1.7 & \tick & \tick & \tick & 0.84 \\
    \code{array\_max\_elim} & 26 & 12 & 1 & 2 & \tick & \tick & \tick & 0.67 \\
    \code{array\_max\_straightforward} & 35 & 8 & 1 & 2  & \tick & \tick & \tick & 0.68 \\
    \code{binary\_search\_seq} & 14 & 8 & 1 & 2 & \tick & \tick & \tick & 0.67\\
    \code{binary\_search\_qp} & 27 & 11 & 1 & 2 & \tick & \tick & \tick & 0.66\\
    \code{dutch\_flag} & 33 & 10 & 1 & 1 & \tick & \tick & \tick & 0.68 \\
    \code{graph\_copy} & 66 & 34 & 2 & 2.7 & \tick & \tick & \tick & 0.84 \\
    \code{graph\_marking} & 82 & 9 & 1 & 2 & \tick & \tick & \tick  & 0.68 \\
    \code{guarded\_by\_monitor} & 51 & 3 & 1 & 2.5 & \tick & \tick & \tick & 0.83\\
    \code{linked\_list} & 211 & 25 & 2 & 2.3 & \tick & \tick & \tick & 7.52\\
    \code{longest\_common\_prefix} & 61 & 4 & 2 & 2.5 & \tick & \tick & \tick & 0.71\\
    \code{sorted\_list} & 23 & 3 & 1 & 3 & \tick & \tick & \tick & 0.70\\
    \end{tabular}
    \caption{Examples that \gaurav{mostly} satisfy the syntactic condition taken from the \Viper{} test suite. For each example, potentially more than one method were inlined. We show the lines of code, lines of annotations for modular verification without errors, seeded errors, average \# spurious errors when verifying modularly without annotations (S.Err.),
    whether inlining is \sound{} (Inl.P.), the \thibault{semantic} condition holds (\thibault{SC}), the structural condition holds (\thibault{Str.C.}) and the average time of 5 runs on a Lenovo T480 with 32 GB, i7-8550U 1.8 GhZ, Windows 10.
    \gaurav{If more than one initial statement was considered in which calls were inlined, then the number of spurious errors is given by the average of spurious errors reported for each of the initial statements.}
    }
    \label{tbl:syntactic_examples}
\end{table}

\clearpage
\section{Examples of Non-Preserving Inlining in Several Verifiers}
\label{app:examples}

\thibault{Our evaluation (\tabref{tbl:results} in \secref{sec:evaluation}) proves the existence of non-preserving inlining
in \grasshopper{}, \nagini{}, \rslviper{}, and \verifast{}.
In this section, we show examples of non-preserving inlining in the other automatic SL verifiers we have mentioned throughout this paper:
\caper{} (\figref{fig:caper}), \refinedc{} (\figref{fig:refinedc}), \steel{} (\figref{fig:steel}),
and \vercors{} (\figref{fig:vercors}).}

\begin{figure}[H]
{
\thibault{
\lstset{emph={requires, function, ensures, assert, region, guards, interpretation, predicate, actions},emphstyle={\bfseries}}
\begin{lstlisting}[numbers=left,language=C]
region A(r, x) {
  guards %G;
  interpretation {
    0 : x |-> 0;
  }
  actions {}
}

function main()
  requires true;
  ensures true;
{
  v := alloc(1);
  [v] := 0;
  callee();
}

function callee()
  requires A(r, x, 0) &*& r@G[1p];
  ensures A(r, x, 0) &*& r@G[1p];
{
  assert r@G[1p];
}   
\end{lstlisting}}
}
\caption{\thibault{Example of non-preserving inlining in \caper{}.
Inlining the call to method \code{callee} on line 15 (\ie replacing it by \code{assert r@G[1p]})
makes verification fail, because this statement asserts the existence of a guard associated with a region,
but \caper{} has not encountered any mention of a region so far, and so it has not created any region
(using the \emph{region creation} proof rule) with the associated guard.
On the other hand, \caper{} is able to successfully verify (modularly) the original program,
with the annotation provided on lines 19 and 20.
The precondition explicitly mentions a region (\code{A(r, x, 0)}), and thus \caper{}
uses the \emph{region creation} proof rule to create this region.
It thus obtains the associated guard, and is able to prove the existence of this guard on line 22.}
}
\label{fig:caper}
\end{figure}

\begin{figure}
\resizebox{1.0\textwidth}{!}{
\thibault{
\begin{lstlisting}[numbers=left,language=C]
#include <stddef.h>
#include <stdbool.h>
#include "refinedc.h"
#include "alloc.h"

typedef struct [[rc::refined_by("runs : {list Z}", "data : {list Z}")]] Sr {
  [[rc::field("&own<array<size_t, {runs `at_type` int size_t}>>")]]
  size_t *runs;
  [[rc::field("{length runs} @ int<size_t>")]]
  size_t runs_length;
  [[rc::field("&own<array<size_t, {data `at_type` int size_t}>>")]]
  size_t *data;
  [[rc::field("{length data} @ int<size_t>")]]
  size_t data_length;
} sr;

[[rc::parameters("p : loc", "q : loc", "runs : {list Z}", "v : Z")]]
[[rc::args("p @ &own<&own<array<size_t, {runs `at_type` int size_t}>>>",
"q @ &own<{length runs} @ int<size_t>>", "v @ int<size_t>")]]
[[rc::ensures("own p : &own<array<size_t, {(runs ++ [v]) `at_type` int size_t}>>")]]
[[rc::ensures("own q : {length (runs ++ [v])} @ int<size_t>")]]
void push_back(size_t **runs, size_t *runs_length, size_t v);

[[rc::returns("{([], []) @ Sr}")]]
sr new_sr();

[[rc::parameters("runs : {list Z}")]]
[[rc::parameters("data : {list Z}")]]
[[rc::parameters("p : loc")]]
[[rc::args("p @ &own<{(runs, data)} @ Sr>")]]
[[rc::ensures("own p : {(runs, data)} @ Sr")]]
void helper(sr *s) { }

[[rc::parameters("data : {list Z}", "l : Z", "p : loc")]]
[[rc::args("p @ &own<array<size_t, {data `at_type` int size_t}>>", "l @ int<size_t>")]]
[[rc::requires("{0 <= l}")]]
[[rc::requires("{l + 1 <= length data}")]]
[[rc::exists("rruns : {list Z}")]]
[[rc::exists("rdata : {list Z}")]]
[[rc::returns("{(rruns, rdata)} @ Sr")]]
[[rc::ensures("own p : array<size_t, {data `at_type` int size_t}>")]]
sr msort(size_t *a, size_t l) {
  sr res = new_sr();
  push_back(&res.data, &res.data_length, a[l]);
  helper(&res);
  return res;
}
\end{lstlisting}}
}
\caption{\thibault{Example of non-preserving inlining in \refinedc{}.
Inlining the call to method \code{helper} on line 45 (\ie removing the call completely, since the body of method \code{helper} is empty)
makes verification fail, because of an incomplete rule (reading a structure from memory does not merge fields whose address has been taken back into the structure).
However, \refinedc{} is able to successfully verify (modularly) the original program,
with the annotation provided on lines 27-31:
This annotation forces \refinedc{} to reverse the effect of taking the address of a field by folding the \code{Sr} predicate.
%
}}
\label{fig:refinedc}
\end{figure}

\begin{figure}
{
\thibault{
\lstset{emph={assume, requires, ensures, module},emphstyle={\bfseries}}
\begin{lstlisting}[language=ML, numbers=left]
module Selectors.Examples

open Steel.Effect.Atomic
open Steel.Effect
open Steel.Reference

assume val assign (#p1 #p2:ref int) (n: int) : Steel unit
  (vptr p1 `star` vptr p2)
  (fun _ -> vptr p1 `star` vptr p2)
  (requires fun _ -> True)
  (ensures fun h0 _ h1 -> n == sel p2 h1)

let callee (r1 r2:ref int) : Steel unit
  (vptr r1 `star` vptr r2)
  (fun _ -> vptr r1 `star` vptr r2)
  (requires fun _ -> True)
  (ensures fun h0 _ h1 -> sel r2 h1 == 5)
  = assign 5

let main (r1 r2:ref int) : Steel unit
  (vptr r2 `star` vptr r1)
  (fun _ -> vptr r2 `star` vptr r1)
  (requires fun _ -> True)
  (ensures fun h0 _ h1 -> sel r2 h1 == 5)
  =
    let _ = get() in
    let x2 = read r2 in
    callee r1 r2
\end{lstlisting}}
}
\caption{\thibault{Example of non-preserving inlining in \steel{}.
The parameters \code{p1} and \code{p2} of library function \code{assign} are existentially quantified.
Inlining the call to method \code{callee} (\ie replacing it by \code{assign 5}) on line 28 makes verification fail:
In this case, the context before the call is $\code{r2} * \code{r1}$ (precondition on line 21), and thus \steel{}
instantiates parameters \code{p1} as \code{r2} and \code{p2} as \code{r1} for the call to method \code{assign},
and \steel{} cannot prove the postcondition.
However, \steel{} is able to successfully verify (modularly) the original program,
with the annotation provided on lines 14-17.
This annotation forces \steel{} to modify the context from $\code{r2} * \code{r1}$ to $\code{r1} * \code{r2}$
before the call to \code{assign}:
In this case, \code{p1} is instantiated as \code{r1} and \code{p2} as \code{r2}, and \steel{} can prove the postcondition.}}
\label{fig:steel}
\end{figure}

\begin{figure}
{
\thibault{
\lstset{emph={}, emphstyle={\bfseries}}
\begin{lstlisting}[numbers=left, language=java]
final class Cell {
  public int value;
  /*@  resource cell() = Perm(value, read);  @*/
}

public class VerCors {
  /*@
  requires Perm(x.value, write);
  ensures Perm(x.value, 1\2) ** x.cell();
  @*/
  static void client(Cell x) { 
    callee(x);
  }

  /*@
  requires Perm(x.value, 1\2);
  ensures x.cell();
  @*/
  static void callee(Cell x) { 
    //@ fold x.cell();
  }
}
\end{lstlisting}}
}
\caption{\thibault{Example of non-preserving inlining in \vercors{}.
The \code{read} permission amount in the definition of resource \code{cell} (line 3) corresponds to
an existentially-quantified positive fractional permission amount.
Inlining the call to method \code{callee} (\ie replacing it by \code{//@ fold x.cell()}) on line 12
makes verification fail: To fold \code{x.cell()}, \vercors{} removes more than half permission to \code{x.value},
and thus it cannot prove the postcondition of method \code{client} on line 9.
On the other hand, \vercors{} is able to successfully verify (modularly) the original program,
with the annotation lines 16-17.
This annotation prevents \vercors{} from removing more than half permission to \code{x.value}.
}}
\label{fig:vercors}
\end{figure}

\end{document}